\documentclass[11pt,a4paper]{article}
\usepackage[T1]{fontenc}
\usepackage[utf8]{inputenc}
\usepackage{lmodern}
\usepackage[english]{babel}
\usepackage{amsmath,amssymb}
\usepackage{graphicx}
\usepackage{xcolor}
\usepackage[margin=25mm]{geometry}
\usepackage{authblk}
\usepackage[font=small,labelfont=bf,labelsep=period]{caption}
\usepackage{microtype}
\usepackage{cite}
\usepackage[colorlinks=true,allcolors=blue!60!black]{hyperref}
\graphicspath{{figures/}}
\usepackage{jabbrv}

\renewcommand{\eqref}[1]{Eq.~(\ref{#1})}
\newcommand{\vc}{\mathbf{c}}
\newcommand{\vx}{\mathbf{x}}
\newcommand{\vJ}{\mathbf{J}}
\newcommand{\vone}{\mathbf{1}}
\newcommand{\Str}{\mathsf{S}}
\newcommand{\Col}{\mathsf{C}}
\newcommand{\Eq}{\mathsf{E}}
\newcommand{\Mom}{\mathsf{M}}
\newcommand{\Proj}{\mathsf{P}}
\newcommand{\Id}{\mathsf{I}}
\newcommand{\Eqf}{\mathcal{E}}
\newcommand{\Rel}{\mathsf{\Omega}}
\DeclareMathOperator{\softmax}{softmax}
\def\splitrange#1 to #2\relax{$#1$ to $#2$}
\newcommand{\caLongEigRangeMath}{\expandafter\splitrange\caLongEigRange\relax}
\newcommand{\eqmain}[1]{Eq.~(#1) of the main text}

\newcommand{\cznOmega}{0.8}
\newcommand{\cznTrialsOne}{2}
\newcommand{\cznTrialsTwo}{10}

\newcommand{\cznRmsOne}{$5.9\times10^{-5}$}
\newcommand{\cznRmsTwo}{$4.5\times10^{-5}$}
\newcommand{\cznRmsLinOne}{0.62}
\newcommand{\cznRmsLinTwo}{0.19}
\newcommand{\cznFvCheck}{$1.2\times10^{-6}$}
\newcommand{\cznSrhOne}{0.00568}
\newcommand{\cznSmeasOne}{0.00568}
\newcommand{\cznSrhTwo}{-0.02032}
\newcommand{\cznSmeasTwo}{-0.02029}
\newcommand{\cznRhoLOne}{0.25}
\newcommand{\cznRhoROne}{0.95}
\newcommand{\cznRhoLTwo}{0.3375}
\newcommand{\cznRhoRTwo}{0.95}
\newcommand{\cznDevSmall}{$1.2\times10^{-5}$}
\newcommand{\cznDevLarge}{$3.0\times10^{-3}$}
\newcommand{\cznSpeedDevSmall}{$2.5\times10^{-5}$}
\newcommand{\cznKmin}{0.0061}
\newcommand{\cznKmax}{0.098}
\newcommand{\cznOmegaDev}{$2.5\times10^{-3}$}
\newcommand{\cznOmegaDevAt}{0.8}
\newcommand{\cznOmegaDevLast}{$8.3\times10^{-4}$}
\newcommand{\cznDRatio}{149}

\newcommand{\cznConsMeas}{0.8521}
\newcommand{\cznConsPred}{0.8521}
\newcommand{\cbRhoC}{0.782}

\newcommand{\cbRhoOne}{0.95}
\newcommand{\cbGammaStar}{0.692}
\newcommand{\cbGammaMeas}{0.692}
\newcommand{\cbDRhoOne}{1.26}

\newcommand{\cbRhoHiStable}{0.7316}
\newcommand{\cbLamStable}{0.79}

\newcommand{\cbAmpDropTwenty}{$2.1\times10^{-3}$}
\newcommand{\cbFloorEnd}{$1.5\times10^{-11}$}
\newcommand{\cbRmsStable}{$1.9\times10^{-4}$}
\newcommand{\cbRmsUnstable}{0.40}
\newcommand{\cbTUnstable}{30}
\newcommand{\cbStepsUnstable}{38}
\newcommand{\cbTStable}{400}
\newcommand{\cbOmegaZero}{1.0}
\newcommand{\cbAlpha}{1.0}
\newcommand{\cbRhoZero}{0.6}
\newcommand{\cbDZero}{0.3769}
\newcommand{\cbStateSpread}{$2.5\times10^{-4}$}

\newcommand{\cbLastStable}{1.99}
\newcommand{\cbFirstDiv}{2.01}
\newcommand{\cbFrozenLast}{0.0019}
\newcommand{\caNX}{64}
\newcommand{\caTwin}{40}
\newcommand{\caKmin}{0.098}

\newcommand{\caShearEig}{-0.597}

\newcommand{\caLongEigRange}{-0.47 to -0.38}
\newcommand{\caGhostMax}{$1.9\times10^{-3}$}
\newcommand{\caAcErr}{$1.3\times10^{-3}$}

\newcommand{\caShErrFar}{$2.1\times10^{-2}$}

\newcommand{\caCons}{$4.4\times10^{-16}$}
\newcommand{\caCtrlEig}{-0.600, -0.600, -0.400, -0.200, -0.100, -0.100}

\newcommand{\ccN}{60}
\newcommand{\ccEdges}{163}

\newcommand{\ccTlong}{300}

\newcommand{\ccParMachine}{1930}
\newcommand{\ccParBase}{2336}
\newcommand{\ccLossMachine}{$3.7\times10^{-6}$}
\newcommand{\ccLossBase}{$3.3\times10^{-4}$}

\newcommand{\ccGamma}{1.305}
\newcommand{\ccOneMinusGamma}{0.305}
\newcommand{\ccFluxFactor}{1.88}

\newcommand{\ccErrMThree}{0.19}

\newcommand{\ccErrBThree}{587}

\newcommand{\ccMassB}{224}
\newcommand{\ccDecayFirst}{0.3054}
\newcommand{\ccImprintOne}{14.3}

\newcommand{\ccImprintHundred}{$8.0\times10^{-5}$}
\newcommand{\ccBaseImprintHundred}{30}
\newcommand{\ccBaseHiddenGrowth}{18.9}
\newcommand{\ccJacMachine}{0.9508}
\newcommand{\ccJacRenorm}{0.9520}
\newcommand{\ccJacClosed}{0.9744}
\newcommand{\ccJacKinCount}{267}
\newcommand{\ccJacDim}{386}

\newcommand{\ccJacBaseAbove}{267}
\newcommand{\ccLawLeadThirty}{3.2}
\newcommand{\ccLawRenThirty}{0.62}

\newcommand{\ccJacMachineHalf}{0.9630}
\newcommand{\ccJacRenormHalf}{0.9626}

\newcommand{\ccAbOneSumMean}{1.0002}
\newcommand{\ccAbOneSumRms}{7.4}

\newcommand{\ccAbOneMassThree}{26}

\newcommand{\ccAbOneErrThree}{5.6}

\newcommand{\ccAbOneClosure}{12 to 14}
\newcommand{\ccAbOneJacMax}{0.9950}

\newcommand{\ccAbTwoLossThreeOverNoChange}{236}

\newcommand{\ccAbTwoErrThreeTwo}{46}

\newcommand{\ccAbTwoKinGrowth}{1.300}

\newcommand{\ccAbTwoUnitCircle}{354}
\newcommand{\ccAbTwoJacDim}{386}

\newcommand{\ccAbTwoFitLaw}{40}
\newcommand{\ccAbThreeDefect}{4.3}
\newcommand{\ccAbThreeClosure}{1.3 to 1.5}
\newcommand{\ccAbThreeErrThree}{0.21}

\newcommand{\ccPersThree}{60}
\newcommand{\caCollapseShear}{0.0083}
\newcommand{\cznOmegaDevTwo}{$2.3\times10^{-3}$}

\newcommand{\cznOmegaDevLastTwo}{$7.8\times10^{-5}$}

\title{The lattice Boltzmann machine}
\author[1]{Hua-Dong Yao\thanks{To whom correspondence should be addressed. E-mail: huadong.yao@chalmers.se}}
\author[2,3]{Sauro Succi}
\affil[1]{Department of Mechanical Engineering, Chalmers University of Technology, Gothenburg, Sweden}
\affil[2]{Fondazione Istituto Italiano di Tecnologia, Center for Life Nano-Neuroscience at La Sapienza, Rome, Italy}
\affil[3]{Department of Physics, Harvard University, Cambridge, Massachusetts 02138, USA}
\date{}

\begin{document}

\maketitle

\begin{abstract}
The lattice Boltzmann (LB) method is widely regarded as a very flexible and efficient solver for a broad class of transport phenomena in fluids, soft matter and related fields. In this paper, we posit that such flexibility is a direct result of its \textit{generative} nature as a rule-driven information-processing machine. Its update, a relaxation toward a local target followed by a shift along the lattice, belongs to the same class of iterated relaxation dynamics as residual networks and transformers, with the difference that in LB the operators are fixed by conservation and symmetry rather than learned from data. We show that three properties of this update are sufficient to promote the emergence of closed macroscopic laws from the underlying iteration process, the physical content of the target deciding the specific law in question. A machine whose target is drawn at random generates the transport law predicted from its moments. A network given the same three properties on a graph, where there is no lattice, is trained on a traffic process and generates a conservation law of its own. The mechanism has a precise limit, where the damping of the non-conserved modes fails, and on the lattice the macroscopic law gives no warning of it. In this reading, LB is a lattice Boltzmann machine whose informing rules are supplied by physics rather than by training. Its three properties then state what a new ingredient should preserve when it enters an LB model, whether it comes from physics or from data.
\end{abstract}

\begin{quote}
\small\textbf{Significance.} The lattice Boltzmann method is widely used to simulate fluids, soft matter and other transport processes. Owing to its flexibility, it has been likened to the deep networks behind generative models, but only by analogy. We show that both belong to one family of algorithms, in which a state is repeatedly propagated and relaxed toward a local target. Networks learn these operators from data, whereas lattice Boltzmann takes them from conservation laws and symmetries. Three basic properties of the lattice Boltzmann algorithm suffice for its iteration to generate a macroscopic law on its own. They stipulate what new physics should preserve, what macroscopic data can teach a learned model and which properties a network needs to generate emergent conservation laws.
\end{quote}


\bigskip

In the last three decades the lattice Boltzmann (LB) method has gained a prominent role in the numerical simulation of complex transport phenomena across a wide range of scales and regimes of motion \cite{Succi2018}. Its update rule consists of two steps: the populations at any given lattice site relax toward a local equilibrium and are then shifted to neighboring sites along the lattice links. The reach of the method now extends well beyond the hydrodynamics it was originally built for \cite{Chen1998,Aidun2010}, encompassing multiphase flow in porous media \cite{Liu2016,Sinha2019}, the dynamics of dense emulsions \cite{Montessori2019,Tiribocchi2025}, turbulence \cite{Chen2003}, relativistic flows \cite{Gabbana2020}, hemodynamics \cite{Bernaschi2009} and, lately, quantum computing as well \cite{Succi2023,Itani2022,Khan2026a}.

LB and deep neural networks have been viewed, by analogy, as instances of a common iterated-relaxation process \cite{Succi2025}. Here we develop this connection by placing them within a single class of iterated relaxation dynamics, with LB distinguished by operators fixed by physics. We show that three properties of the update are sufficient for the iteration to generate a closed macroscopic law: conservation of a set of linear functionals, a local-equilibrium target that acts as a projector, and a spectral gap of the step composed of relaxation and shift. The equilibrium projection preserves a small set of conserved quantities and depends only on them, while the composed step damps the remaining modes. In this sense, LB is generative, since the macroscopic law emerges from the repeated application of a local rule operating in a higher-dimensional (phase) space, a view anticipated by Succi \cite{Succi2015}, who described LB as a kinetic generator of partial differential equations. We use the term ``generative'' with this meaning throughout. The lattice Boltzmann machine is the same dynamical system, yet viewed as an information-processing machine that generates the macroscopic law.

The update is constructed top-down, following Higuera, Succi and Benzi \cite{Higuera1989}, who built the collision operator from the conservation of mass and momentum %
and the lattice symmetry, with eigenvalues chosen independently of the collision rules of a lattice gas. The lattice Bhatnagar--Gross--Krook (BGK) scheme of Qian, d'Humi\`eres and Lallemand \cite{Qian1992} reproduces the weakly compressible
Navier--Stokes equations with a single relaxation time.
As mentioned above, LB does not directly discretize these equations: it lifts the hydrodynamic fields to the phase space of the populations, which carries a discrete velocity index in addition to position. At this level, advection and diffusion emerge from the underlying application of a linear shift at constant velocity and a collisional relaxation to a nonlinear local equilibrium. This kinetic lifting introduces extra variables but in return it equips the method with the structure of a machine, i.e., a fixed linear mixing composed with a local nonlinear map.

Interestingly, the same distinction appears in quantum mechanics. In a suitable representation, the Dirac equation has the structure of the machine, where a propagation of the spinor components along the light cone is composed with a local rotation associated with the mass term. This composition underlies the quantum lattice Boltzmann scheme of Succi and Benzi \cite{Succi1993}. The Schr\"odinger equation, which arises from the Dirac equation in the nonrelativistic limit, does not have this structure in its usual formulation, since its kinetic term couples the wave function across space. The Dirac step is unitary and has no dissipative gap, so it shares the mixing and the local map with LB but not its dissipative relaxation. The machine structure belongs to the kinetic level and is not explicit in the macroscopic form.

Much of the reach of LB has come from modifying the collision or adding interactions, while leaving the basic propagation step unchanged, even though extending the propagation to larger neighborhoods (high-order lattices) has proved very effective in probing deep nonequilibrium phenomena \cite{Shan2006,Montessori2015,Xu2024}. The multiple-relaxation-time collision of d'Humi\`eres \cite{dHumieres1992} replaced the single relaxation rate with one rate per kinetic mode. The pseudopotential interaction of Shan and Chen \cite{Shan1993} added a density-dependent force between neighboring sites and produced nonideal fluids and phase separation within the same update. Entropic and state-dependent relaxation made the rate a function of the local state \cite{Karlin1998,Ansumali2002,Karlin2014}. The macroscopic equations recovered by such a scheme were derived for a general equilibrium through the equivalent-equation analysis, and they depend on the chosen equilibrium \cite{Dubois2008}.

The question of which kinetic models can reproduce a prescribed conservation law has been studied through relaxation and discrete kinetic schemes. Jin and Xin \cite{Jin1995} approximated a conservation law by a linear hyperbolic system with relaxation toward the prescribed flux, and the system recovers the law with a diffusive correction under the sub-characteristic condition of Liu \cite{Liu1987}. Bouchut \cite{Bouchut1999} and Aregba-Driollet and Natalini \cite{AregbaDriollet2000} constructed discrete kinetic models of BGK type whose equilibria generate a prescribed system of conservation laws. In that theory the law is given and the kinetic model is built to reproduce it. Related invariant-manifold analyses established conditions for second-order hydrodynamic accuracy in lattice BGK dynamics \cite{Brownlee2007} and constructed an exact spectral closure of the linear BGK equation \cite{Kogelbauer2025}.

Deep neural networks have also been viewed as discrete dynamical systems in a related sense. Residual networks were interpreted as discretized dynamical systems, which led to continuous-depth models \cite{He2016,E2017,Haber2018,Chen2018}. Succi \cite{Succi2025} expressed feed-forward networks in relaxation form, with each layer relaxing toward a local target, strongly reminiscent of the LB collisional update. Transformers, the architecture behind large language models, add state-dependent mixing through attention \cite{Vaswani2017}, and neural integro-differential equations provide a related framework for learning nonlocal dynamics \cite{Zappala2023}.

In the opposite direction, neural networks have been incorporated into kinetic schemes. They have served as learned collision operators under conservation and lattice-symmetry constraints \cite{Corbetta2023,Ortali2024}, as networks that set the relaxation rates of the non-conserved moments from the local state \cite{Bedrunka2024}, and as a learned equilibrium inside a solver that preserves transport, relaxation and moment conservation \cite{LaraBenitez2025}. Differentiable LB implementations make such components trainable through the lattice update itself \cite{Ataei2024,Bedrunka2021}.

Despite this two-way exchange, these two readings of LB have not been explicitly connected. The numerical reading treats the macroscopic law as an input and the kinetic update as a means of reproducing it. It reflects the success of the method as a simulation tool and is formalized in the relaxation-scheme theory. The structural reading has not established which properties of the update would let a network generate a closed macroscopic law. To our knowledge, this connection has not been tested directly.

In this paper, we test our statement in four concrete cases. First, we ask a machine with a random constrained target to generate the transport law predicted from its moments. Second, we drive the same machine to the point where its composed step loses stability, to examine where the mechanism stops and whether the breakdown is visible in the macroscopic law. Third, we learn a collision operator through the machine from macroscopic fields, to determine which of its eigenvalues can be identified from the data. Finally, we give a network the three properties on a graph, where there is no lattice, and train it on a traffic process, to test whether it generates its own law and clarify the role of each property separately.

\section*{Results}

If the macroscopic law emerges from the iteration, then a target designed without physics should still generate a law, and the law should fail where the damping of the non-conserved modes fails. Data about the law should then determine only the part of the machine that produces it, and the properties rather than the lattice should play the key role.

\subsection*{The lattice Boltzmann machine and three properties}

We start from the standard LB update. The populations $f_i(\vx,t)$, $i=1,\dots,Q$, give the amount of fluid at lattice site $\vx$ and time $t$ moving with discrete velocity $\vc_i$. With unit lattice spacing and time step, the single-relaxation-time update reads
\begin{equation}
\label{eq:lbgk}
f_i(\vx+\vc_i,t+1)=f_i(\vx,t)-\omega\bigl[f_i(\vx,t)-\Eq_i(f(\vx,t))\bigr],
\end{equation}
where $0<\omega<2$ is the relaxation rate, and $\Eq$ maps the populations at a site to their equilibrium. Writing $f_t$ for the population field, we factor the step into a local collision followed by the shift,
\begin{equation}
\label{eq:factor}
\begin{gathered}
f_{t+1}=\Str^{-1}\Col(f_t),\qquad
\Col(f)=(1-\omega)f+\omega\,\Eq(f),\\
(\Str^{-1}f)_i(\vx)=f_i(\vx-\vc_i),
\end{gathered}
\end{equation}
where $\Col$ is the local collision operator relaxing the populations toward equilibrium, $\Str^{-1}$ is a parameter-free linear operator for the shift, and $\Eq$ is the equilibrium operator. %

\begin{figure}[tbp]
\centering
\includegraphics[width=\textwidth]{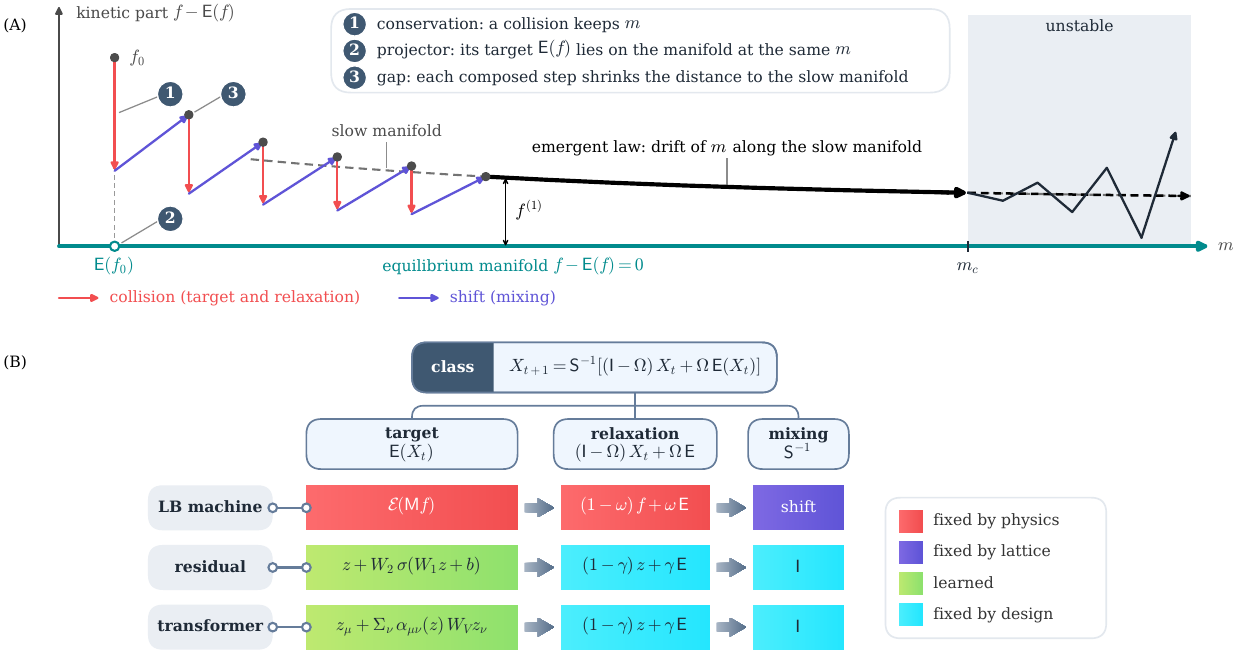}
\caption{The lattice Boltzmann machine, the mechanism, and the class. (A) The mechanism, drawn in the kinetic part $f-\Eq(f)$ against the conserved moments $m$, in which the equilibrium manifold is the axis of $m$. The slow manifold (the dashed line) is offset from the equilibrium manifold by the first-order correction $f^{(1)}$. The drift of $m$ (the black arrow) indicates the emergent law. Stability is lost past $m_c$ ($\rho_c$ of the scalar machine) in the shaded region. The numbers mark where the three properties act.
(B) One step of the class, of the LB machine, of a residual block and of a transformer block, as the composition of a target, a relaxation toward it and a linear mixing.}

\label{fig:class}
\end{figure}

The conserved moments are collected by a $(d+1)\times Q$ matrix $\Mom$, where $d$ is the spatial dimension, so that $\Mom f=(\rho,\vJ)$, with $\rho$ and $\vJ$ denoting the density and the current. The equilibrium depends on the populations only through these moments and reproduces them,
\begin{equation}
\label{eq:eqmom}
\Eq=\Eqf\circ\Mom,\qquad \Mom\,\Eqf(m)=m,
\end{equation}
where $\Eqf$ maps the moments to equilibrium populations. The form of $\Eq$ establishes the first two of three properties, namely conservation and projector target. %

The conservation follows directly from the moment-reproduction condition in \eqref{eq:eqmom}. Since $\Mom\Eq=\Mom$, the collision preserves the conserved moments locally as $\Mom\Col(f)=\Mom f$ at every site, and the shift conserves their sums over the lattice.

The target is a projector under the same condition. %
We get $\Eq[\Eq(f)]=\Eq(f)$, and its Jacobian $\Proj(f)=D\Eqf(\Mom f)\,\Mom$ satisfying $\Proj^2=\Proj$ (Supplementary Material, section S1). Then, the collision Jacobian is
\begin{equation}
\label{eq:proj}
D\Col(f)=\Proj(f)+(1-\omega)\bigl[\Id-\Proj(f)\bigr],
\end{equation}
with eigenvalue $1$ on the range of $\Proj(f)$, the tangent space of the equilibrium manifold, of dimension $d+1$, and eigenvalue $1-\omega$ on the kinetic subspace $\ker\Mom$, of dimension $Q-d-1$. This decomposition holds at every admissible state and not only at equilibrium. The entropic equilibrium, which maximizes a discrete entropy at fixed conserved moments \cite{Karlin1998}, is a target of this kind. The same splitting into conserved and relaxing modes organizes the Carleman-linearized collision into blocks for quantum simulation \cite{Yao2026a,Yao2026b}.

The spectral gap concerns the composed collision-shift step. For $0<\omega<2$ the kinetic eigenvalue satisfies $|1-\omega|<1$, so kinetic perturbations decay at every collision while the conserved ones persist. This local damping, however, does not by itself guarantee a gap for the complete step. Linearized about a uniform state, one step at wavenumber $\mathbf k$ is a growth factor $A(\mathbf k)$ (Supplementary Material, section S1). Its eigenvalues continuously connected to one at $\mathbf k=0$ are the hydrodynamic branches, and its remaining eigenvalues are the kinetic branches (Supplementary Material, Fig.~S2). The third property requires stability of the composed step and separation of the kinetic and hydrodynamic branches at long wavelengths. Apart from the conserved eigenvalues at $\mathbf k=0$, the eigenvalues of $A(\mathbf k)$ need to have modulus below one at long wavelengths and no larger than one at any wavenumber, so that the iteration can settle on a slow manifold near the equilibrium manifold (Fig.~\ref{fig:class}(A)). For the standard equilibrium this is part of the linear stability condition of the scheme. For a general target it is the condition that the shift does not undo the contraction of the collision.
A multiple-relaxation-time collision \cite{dHumieres1992} replaces $\omega$ by a matrix with one rate per kinetic mode and keeps the first two properties, with kinetic eigenvalues $1-\omega_k$.

The kinetic subspace can be divided into two parts. The transport directions are the kinetic directions contributing to the fluxes of the conserved moments, the current for a scalar machine and the stress for a fluid. The ghost directions are the remaining directions without contributions, in the sense of the non-hydrodynamic modes \cite{Dellar2002}.  Because the split is defined through the fluxes rather than through specific lattice moments, it can extend beyond lattice-based updates wherever the conserved fluxes are defined. The state then has sectors: conserved, transport and ghost (CTG). With a single rate, the two kinetic sectors share the eigenvalue $1-\omega$, and the split is structural only. Separate relaxation rates introduce a spectral distinction, as in the multiple-relaxation-time reference. %

\subsection*{The common class of iterated relaxation dynamics}

Written as a collision followed by a shift, the LB update combines relaxation toward a local target with a fixed linear operator. Deep networks can also be expressed in a similar form. A residual block updates a feature vector as $z_{l+1}=z_l+\gamma\,\phi_l(z_l)$, with $\phi_l$ a small network such as $W_2\sigma(W_1z+b)$ \cite{He2016}. This is equivalent to relaxation at the rate $\gamma$ toward the target $z+\phi_l(z)$, where $\gamma=1$ in the standard block and may be learned in some variants. The layer index then serves as the time index, and each layer relaxes toward a layer-dependent target \cite{Succi2025}. Similarly, a transformer block has attention operation inside the target \cite{Vaswani2017}, which mixes the state across positions with state-dependent coefficients (Supplementary Material, section S1).

\begin{figure}[tbp]
\centering
\includegraphics[width=\textwidth]{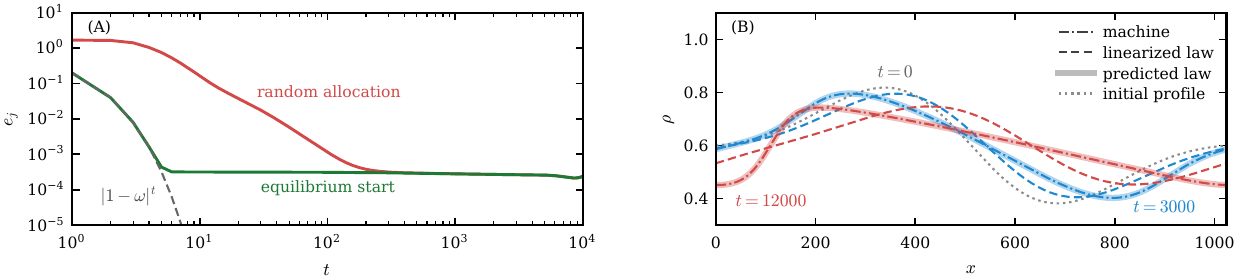}
\caption{The scalar machine and the law it generates. (A) Deviation $e_j$ of the kinetic current from its first-order value for the first random machine at $\omega=\cznOmega$, started at equilibrium and from a random allocation of the same density, with $|1-\omega|^t$ (dashed). (B) Density of the same machine at two times (dash-dotted) against the solution of the law linearized about the mean density (dashed) and of the predicted emergent law (solid thick), with the initial profile (dotted).}
\label{fig:emerge}
\end{figure}

The three dynamics therefore share a common one-step structure of the class, as illustrated in Fig.~\ref{fig:class}(B).
We consider dynamics on a state $X_t$ that advance by one relaxation and one fixed linear operator,
\begin{equation}
\label{eq:class}
X_{t+1}=\Str^{-1}\bigl[(\Id-\Rel)\,X_t+\Rel\,\Eq(X_t)\bigr],
\end{equation}
where $\Str^{-1}$ is a linear operator for the shift that carries no adjustable parameter, $\Eq$ is a nonlinear map of the state to a target, and $\Rel$ is a relaxation operator whose eigenvalues lie in $(0,2)$. In this range, the relaxation moves the state toward the target. The LB update \eqref{eq:factor} is \eqref{eq:class} with $\Str^{-1}$ along the lattice links, $\Rel=\omega\Id$ and $\Eq$ the local equilibrium. The residual block is \eqref{eq:class} with $\Str^{-1}=\Id$, $\Rel=\gamma\Id$ and $\Eq(z)=z+\phi_l(z)$.
In a network $\Eq$ contains learned matrices and $\Str$ is trivial, whereas in LB $\Eq$ is fixed by conservation and symmetry and $\Str$ by the lattice associated. %
The three properties of LB distinguish its step from that of a network.

\subsection*{Emergence as generative mechanism}

The Chapman--Enskog expansion is used to analyze how the three properties lead to a closed macroscopic law at long wavelengths, where the populations stay close to local equilibrium, and for a smooth target.
We denote the conserved moments as $m=\Mom f$, and let $\Eq=\Eqf\circ\Mom$ satisfy \eqref{eq:eqmom} with constant $\omega$. %
Furthermore, define the flux $\Phi^\beta(m)$ and the second-moment tensor $\Psi^{\beta\gamma}(m)$ of the target as the first and second velocity moments of $\Eqf(m)$ projected on the conserved moments by $\Mom$ (Supplementary Material, section S1).
The current components involve velocity moments one order higher than the density components.
A multiscale expansion of \eqref{eq:lbgk} in a small parameter $\epsilon$ gives the law at the first order on the slow time $t_1$ as $\partial_{t_1}m+\partial_\beta\Phi^\beta=0$,
and at the second order as
\begin{equation}
\label{eq:ce}
\begin{split}
\partial_t m+\partial_\beta\Phi^\beta(m)&=
\Bigl(\frac1\omega-\frac12\Bigr)\partial_\beta\Bigl[\partial_\gamma\Psi^{\beta\gamma}(m)+\partial_{t_1}\Phi^\beta(m)\Bigr]\\
&\quad+O(\epsilon^3),
\end{split}
\end{equation}
with $\partial_{t_1}\Phi^\beta=-D\Phi^\beta(m)\,\partial_\gamma\Phi^\gamma$ derived from the first-order law. %

\eqref{eq:ce} is closed in $m$. Its form, a system of conservation laws with a second-order correction, is set by the moments $\Phi$ and $\Psi$ of the target, and its coefficient by the kinetic eigenvalue through $1/\omega-1/2$. With separate rates, the coefficient is set by the transport rate alone, since the first-order correction $f^{(1)}$ to the populations enters the law only through the fluxes, and the ghost rate enters the slow (hydrodynamic) eigenvalue no earlier than the third order in the wavenumber (Supplementary Material, section S1).
Conservation makes $m$ a slow variable. The projector form makes $\Eqf(m)$ the leading term and places $f^{(1)}$ in the kinetic subspace by solvability. The gap makes the kinetic part relax faster than $m$ changes, which orders the expansion. For a lattice approximation to the Maxwell--Boltzmann equilibrium, $\Phi$ gives the Euler fluxes and $\Psi$ supplies the moments needed for the viscous correction. Thus, \eqref{eq:ce} is the weakly compressible Navier--Stokes system with kinematic viscosity $c_s^2(1/\omega-1/2)$ \cite{Succi2018}, where $c_s$ is the lattice sound speed. %

To remove the physics from the target, we build a D1Q3 machine (one dimension, three velocities) with $c_i\in\{-1,0,1\}$ and a single conserved moment given by the density $\rho=\sum_if_i$. Its target is an arbitrary smooth map $\rho\mapsto(\Eqf_{-1},\Eqf_0,\Eqf_{+1})$ with $\sum_i\Eqf_i=\rho$, specified by its first two moments,
\begin{equation}
\label{eq:FPi}
F(\rho)=\Eqf_{+1}-\Eqf_{-1},\qquad \Pi(\rho)=\Eqf_{+1}+\Eqf_{-1},
\end{equation}
which are set to random cubic polynomials of $\rho \in [0.2,1]$. A sampled target is kept when the target is admissible, the transport slow and diffusive, and the composed step passes the stability test. For the two accepted targets, every component of the target increases with density, which ensures linear stability at every wavenumber throughout the density range (Supplementary Material, sections S1 and S2).
The target specification does not involve any isotropy, Hermite structure or physical model. For this machine, \eqref{eq:ce} reduces to a nonlinear conservation law with a nonlinear diffusion,
\begin{equation}
\label{eq:P}
\begin{gathered}
\partial_t\rho+\partial_xF(\rho)=\partial_x\bigl[D(\rho)\,\partial_x\rho\bigr],\\
D(\rho)=\Bigl(\frac1\omega-\frac12\Bigr)\bigl[\Pi'(\rho)-F'(\rho)^2\bigr],
\end{gathered}
\end{equation}
where the prime denotes the derivative with respect to $\rho$. The positivity of $D$ is the sub-characteristic condition of the kinetic model \cite{Liu1987,Jin1995}.
The resulting law has the form that relaxation-scheme theory assigns to a scalar kinetic model \cite{Jin1995,AregbaDriollet2000}. %
\eqref{eq:P} is our prediction, and specifies the flux function with which the machine transports, the diffusion it adds, and the Rankine--Hugoniot speed $U=[F(\rho_L)-F(\rho_R)]/(\rho_L-\rho_R)$ at which a front between two densities travels. The quantities are determined by the two random functions. %

We run two such machines at $\omega=\cznOmega$ (Supplementary Material, section S2). %
The first machine transports forward at low density and backward at high density, whereas the second transports backward throughout. The diffusivity in the respective machines varies by a factor of 1.7 and 2.1 over the density range.

\begin{figure}[tbp]
\centering
\includegraphics[width=\textwidth]{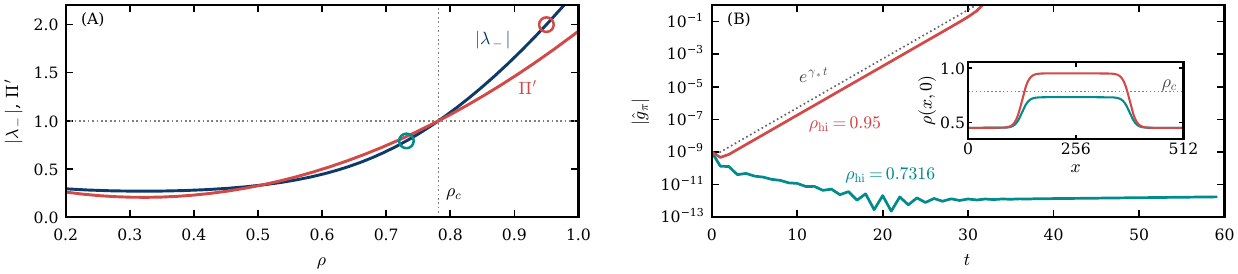}
\caption{Where the mechanism stops. (A) Modulus of the checkerboard eigenvalue $\lambda_-$ of the composed step and $\Pi'$ against density for the random target rejected by the stability test, at $\omega=\cznOmega$. The dotted lines mark $\rho_c$ and the level one, and the circles mark the plateau densities of the two runs. (B) Amplitude $|\hat g_\pi|$ of the checkerboard mode against time for the two runs, with the predicted growth $e^{\gamma_* t}$ (dotted). The mode is measured on the plateaus of the two initial profiles shown in the inset.}
\label{fig:limit}
\end{figure}

Before the law can describe the density evolution, the kinetic part of the populations must relax to a state determined by the density field and its gradients. The expansion specifies this dependence. To first order, the kinetic current $j=\sum_i c_i g_i$, which measures the transport component of the departure $g=f-\Eq(f)$, is slaved to the density gradient,
\begin{equation}
\label{eq:j1}
\begin{gathered}
j^{(1)}=-\frac{1}{\omega}\bigl[\Pi'(\rho)-F'(\rho)^2\bigr]\partial_x\rho,\\
\Bigl(1-\frac{\omega}{2}\Bigr)\,j^{(1)}=-D(\rho)\,\partial_x\rho,
\end{gathered}
\end{equation}
where $j^{(1)}$ is the current of $f^{(1)}$. The second relation identifies the diffusive flux in \eqref{eq:P} as the average of this slaved current before and after collision, since the collision multiplies the kinetic current by $1-\omega$. The emergent law can therefore be tracked at the kinetic level through the relative deviation $e_j=\|j-j^{(1)}\|/\|j^{(1)}\|$ of the measured current from its slaved value, where $\|\cdot\|$ denotes the root-mean-square norm over the lattice.

Fig.~\ref{fig:emerge}(A) shows $e_j$ for the first machine. %
The populations start either at equilibrium or from a random allocation of the same density among the three populations at each site, as illustrated by $f_0$ in Fig.~\ref{fig:class}(A). The equilibrium start has $j=0$ and thus an initial deviation $e_j=1$. Its deviation decays as $|1-\omega|^t$ at the rate set by the kinetic eigenvalue of the collision, and reaches about $3\times10^{-4}$ after six steps. The deviation then remains at this level throughout the $10^4$ steps, over which the density profile evolves under the law (Fig.~\ref{fig:emerge}(B)). The random allocation starts with a deviation more than two orders of magnitude larger. Its first shift converts part of the random allocation into lattice-scale density variations, for which the small-wavenumber slaving relation does not apply. Its deviation therefore decays more slowly and approaches that of the equilibrium start after about 250 steps, once these variations have been smoothed out. The kinetic current thus approaches its slaved value before the macroscopic profile evolves appreciably, with the random start requiring additional time for the lattice-scale variations to decay. This separation of time scales is the emergence mechanism in the machine. %

The linear response provides a direct test of \eqref{eq:P}. A sinusoidal perturbation of wavenumber $k$ about $\rho_0$ should travel at the phase speed $F'(\rho_0)$ and decay at the rate $\Gamma=D(\rho_0)k^2$. For both machines and two base densities at the smallest wavenumber, the phase speed agrees to \cznSpeedDevSmall{} in relative terms, and the decay rate with the emergent law to \cznDevSmall{}. At larger wavenumber, the decay rate follows the exact lattice growth factor $A(k)$, which is the finite-wavenumber correction neglected by the expansion (Supplementary Material, Fig.~S1(C) and (D)). The diffusivity also carries a specific dependence on the kinetic relaxation. Over a scan of the relaxation rate up to $\omega=1.98$, the measured decay rate of one mode of the first machine follows $(1/\omega-1/2)[\Pi'-F'^2]$ to \cznOmegaDevLast{} at $\omega=1.98$, where the diffusivity has fallen by two orders of magnitude (Supplementary Material, Fig.~S1(B)). The emergent law thus captures the small-wavenumber limit of the linear response of the composed step, including its dependence on the kinetic eigenvalue $1-\omega$, as predicted by the derivation.

Agreement in the linear regime does not, however, establish whether the law holds when the density varies over a wide range and the flux is strongly nonlinear. To test this, we initialize both machines with a smooth profile spanning the middle half of the density range on a periodic lattice and solve \eqref{eq:P} from the same profile using a conservative finite-volume method (Supplementary Material, section S2). Fig.~\ref{fig:emerge}(B) shows the first machine. Its profile steepens on one side and spreads on the other, consistent with the density dependence of $F'$. It follows the predicted emergent law with a final-time root-mean-square deviation of \cznRmsOne{} normalized by the standard deviation of the profile. By contrast, the law linearized about the mean density, with constant $F'$ and $D$, fails to reproduce this evolution (dashed curves). The agreement therefore supports the nonlinear law rather than its constant-coefficient approximation. A front connecting two densities that admit a single viscous shock travels at the Rankine--Hugoniot speed determined by the random flux, and the second machine likewise follows its own predicted law and front speed.

Therefore, a random target with the three properties %
generates the nonlinear transport law %
predicted from its two moments,
with the coefficient set by the kinetic eigenvalue. The law is a property of the iterated population field as a whole, in the sense that macroscopic behavior emerges from microscopic dynamics in nonequilibrium systems \cite{Nicolis1977}.
However, the derivation is a small-wavenumber statement, and the next section examines where it stops.

\begin{figure}[tbp]
\centering
\includegraphics[width=\textwidth]{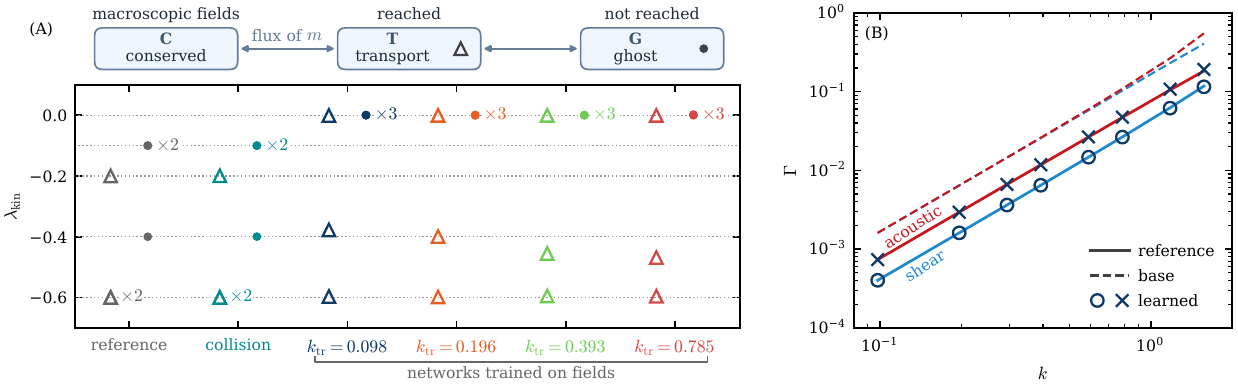}
\caption{A collision learned through the machine from density and current fields. (A) Kinetic eigenvalues $\lambda_{\rm kin}$ of the collision Jacobian at equilibrium in the transport (triangles) and ghost (dots) sectors, for the reference, the control network trained on the collision, and the networks trained through the machine on density and current fields at four training wavenumbers $k_{\rm tr}$.  The number of coinciding eigenvalues is shown beside the markers. %
(B) Damping rates $\Gamma$ of the shear and acoustic branches of the linearized step against $k$ for the reference (lines), the BGK machine at $\omega=1$ (dashed) and the machine trained at $k_{\rm tr}=\caKmin$ (markers).}
\label{fig:caseA}
\end{figure}

\subsection*{Where the mechanism stops}

The third property concerns both long-wave scale separation and stability of the composed step.
The checkerboard channel of the scalar machine allows the latter to be examined exactly. At the wavenumber $k=\pi$, the shift reverses the sign of the Fourier amplitudes of the two moving populations, and the three eigenvalues of $A(\pi)$ are $\omega-1$ and a pair $\lambda_\pm$ that depends on $\omega$ and $\Pi'$ at the uniform density (Supplementary Material, section S1). The modulus of $\lambda_-$ reaches one at $\Pi'=1$ regardless of $\omega$, and exceeds one for $\Pi'>1$.
The quantity $\rho-\Pi$ is the remaining target population associated with $c_i=0$. The composed step therefore reaches its stability limit when this population stops increasing with density, and beyond this threshold the checkerboard mode is amplified at every step. The standard equilibrium has $\Pi'=1/3$ and never reaches the threshold. In the CTG decomposition, the transport sector mediates the emergent law at hydrodynamic scales, whereas at $k=\pi$ the instability arises through the shift-induced coupling of the conserved and ghost sectors, with the transport sector decoupled (Supplementary Material, section S1). %

A random target that fails the composed-step test of the previous section reaches $\Pi'=1$ at $\rho_c=\cbRhoC$ (Fig.~\ref{fig:limit}(A)). At $\rho=\cbRhoOne$ its most amplified mode is $k=\pi$ (Supplementary Material, Fig.~S2), with a per-step growth rate $\gamma_*=\ln|\lambda_-|=\cbGammaStar$, while the law \eqref{eq:P} remains regular with $D>0$. We therefore predict that, on a plateau at $\rho=\cbRhoOne>\rho_c$, the lattice-scale checkerboard mode grows at rate $\gamma_*$ and drives the dynamics away from the slow manifold, even though the predicted macroscopic law stays smooth and well defined.

To test this prediction, we run the machine at $\omega=\cznOmega$ from two profiles of the same shape (the inset of Fig.~\ref{fig:limit}(B)), one with a plateau below $\rho_c$ and the other with a plateau at $\cbRhoOne$. Both profiles are seeded with a small checkerboard perturbation along the ghost direction (Supplementary Material, section S2). Fig.~\ref{fig:limit}(B) shows the amplitude $|\hat g_\pi|$ of that mode on the plateau, defined by the Fourier component at $k=\pi$ of the kinetic departure from equilibrium, $g=f-\Eq(f)$. Below $\rho_c$, the amplitude decays within twenty steps, and the density follows the solution of \eqref{eq:P} (Supplementary Material, section S2). Above $\rho_c$, the amplitude grows at \cbGammaMeas{} per step against the predicted \cbGammaStar. The density develops a two-site pattern on the plateau, and at $t=\cbTUnstable$ its root-mean-square deviation from the smooth solution of \eqref{eq:P} reaches \cbRmsUnstable{} of the standard deviation of the profile, although that solution shows no sign of instability. The machine diverges a few steps later. The finite-wavenumber instability thus marks the breakdown of the emergent dynamics, while the macroscopic law derived from the small-wavenumber limit cannot detect it. This is why the third property should hold for the composed step.

\subsection*{Learning collision through the machine}

When the machine generates the macroscopic law, the inverse problem is which parts of the collision can be determined from macroscopic fields. As in networks whose outputs leave some parameters unconstrained \cite{Ganguly2026}, the LB collision has kinetic rates that do not enter the hydrodynamic law, and the transport–ghost split identifies the ghost directions in advance. We test this by training a collision correction from density and current fields alone, with the shift included in each iteration.

Our reference is a D2Q9 multiple-relaxation-time machine (two dimensions, nine velocities) with the BGK equilibrium and one relaxation rate per kinetic moment. Its moment basis comprises conserved density and current and six kinetic moments: energy, squared energy, two energy-flux components, normal stress, and shear stress (Supplementary Material, section S3). Its six kinetic eigenvalues are $-0.2$, $-0.4$, $-0.1$ (twice) and $-0.6$ (twice). The learned machine starts from the BGK machine at $\omega=1$, whose kinetic eigenvalues all vanish, and adds a projected correction,
\begin{equation}
\label{eq:learned}
\begin{gathered}
\Col_\theta(f)=\Col(f)+\bigl[\Id-\Proj(f)\bigr]\,r_\theta(f),\\
r_\theta(f)=\hat r_\theta(f)-\hat r_\theta[\Eq(f)],
\end{gathered}
\end{equation}
with $\hat r_\theta$ a multilayer perceptron of the local populations. The projection ensures that the correction conserves density and current for any network output, since $\Mom(\Id-\Proj)=0$. The subtraction keeps the equilibrium a fixed point. No further constraints are imposed. The fields are uniform in one lattice direction, allowing both shear and longitudinal waves of wavenumber $k$ along the other. We train the correction to match the reference density and current over short windows, using random low-Mach initial fields at a single training wavenumber $k_{\rm tr}$. The loss is differentiated through every collision and shift. The network never sees the reference populations.

The macroscopic law in this geometry has damping coefficients from the shear viscosity of the transverse wave and the longitudinal viscosity of the acoustic wave. The reference realizes the latter through the relaxation channels of the energy and normal-stress rates. While isotropy ties the normal- and shear-stress rates in the reference, the learned correction imposes no such tie. At a single training wavenumber, the shear wave determines the shear rate, whereas acoustic damping constrains only the combination of the energy and normal-stress rates.
Starting from the BGK machine, in which all six kinetic eigenvalues vanish, we expect training to move only two of them appreciably away from zero: one toward the reference shear eigenvalue, and the other toward an effective longitudinal value that reproduces the acoustic damping without matching any reference eigenvalue. We expect the remaining four to stay near zero.

Fig.~\ref{fig:caseA}(A) shows the kinetic eigenvalues $\lambda_{\rm kin}$ of the collision Jacobian at equilibrium. %
The control network trained on the collision recovers all reference eigenvalues. In contrast, the networks trained on fields move only two eigenvalues away from zero.
These are the shear eigenvalue at $\caShearEig$ and an effective eigenvalue from \caLongEigRangeMath{} of the mode combining energy and normal-stress channels.
The same two channels also form a second mode, but waves along a single lattice direction constrain only the combination that enters acoustic damping. The training loss is therefore nearly insensitive to this second mode, whose eigenvalue stays within \caGhostMax{} of its initial value of zero in the BGK machine at $\omega=1$.
Those eigenvalues of the squared energy and two energy-flux modes in the ghost sector also lie within the same range, as their relaxation rates do not enter the derived macroscopic law. %
Fig.~\ref{fig:caseA}(B) shows the damping rates of the law for the field-trained network at $k_{\rm tr}=\caKmin$. %
After the training, the rates of the shear and acoustic branches agree with the reference. Since the learned eigenvalues depend on $k_{\rm tr}$ (Fig.~\ref{fig:caseA}(A)), the agreement degrades in proportion to $|k^2-k_{\rm tr}^2|$ (Supplementary Material, Fig.~S3).

The data fix the hydrodynamic branches at the sampled wavenumber through combinations of the kinetic rates, while eigenvalues with little influence on those branches remain weakly constrained. In the language of networks, the function constrains the weights to the extent that they enter the emergent law, and the spectrum identifies the relevant kinetic directions in advance.

\begin{figure}[tbp]
\centering
\includegraphics[width=\textwidth]{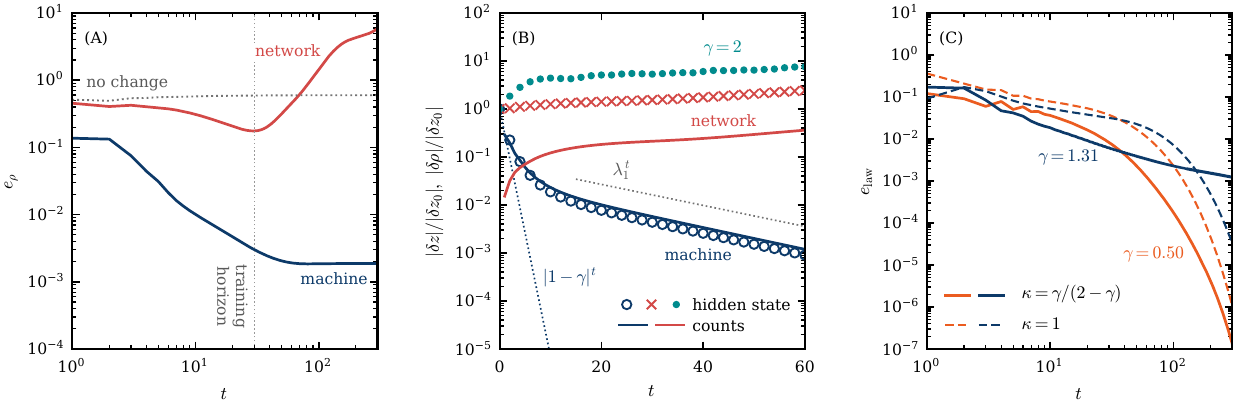}
\caption{The machine on a graph. (A) Held-out error $e_\rho$ against rollout length for the graph machine, the message-passing network and the no-change reference (dotted), with the training horizon as a vertical line. (B) Difference between two runs started from different allocations of the same counts, for the machine (kinetic difference, circles; imprint on the counts, solid), for the message-passing network perturbed in its hidden channels (crosses and solid) and for the machine trained with $\gamma$ fixed at $2.0$ (dots), with $|1-\gamma|^t$ and $\lambda_1^t$ as dotted lines. (C) Error $e_{\rm law}$ of the closed law in the counts alone against the machine, for the learned $\gamma$ and for $\gamma$ fixed at $0.5$, with $\kappa=1$ (dashed) and $\kappa=\gamma/(2-\gamma)$ (solid).}
\label{fig:caseC}
\end{figure}

\subsection*{Three properties tested on a graph}

In LB, the three properties are realized through a lattice-based construction. We test whether they are sufficient without a lattice by imposing them directly on a network defined on a graph. The network has no other LB-specific structure and is trained on a process with no kinetic variables. If the three properties are sufficient, the network should develop a slow manifold and an emergent closed law. Changing each construction element then tests which part of that behavior depends on it.

The process describes a conserved number of vehicles moving on a random road-like graph with $\ccN$ nodes and $\ccEdges$ bidirectional edges under a congestion rule of the cell-transmission type \cite{Daganzo1994}. The flow from $v$ to $w$ grows with the count at $v$, is throttled by the count at $w$ and is bounded by the link capacity $\chi_{vw}$ (Supplementary Material, section S4). The process is nonlinear, but its state consists of the node counts. It relaxes toward a fixed point. The learners observe only the node counts $\rho_v(t)$ over short windows of training trajectories, and are evaluated on held-out trajectories.

Our constrained learner, the graph machine, is a phase-space network on the graph. At each node $v$, its state contains one variable $z(v,r)$ for each outgoing edge $r$ and one rest variable $z(v,0)$. The node count is the sum over all slots, $\rho_v=z(v,0)+\sum_r z(v,r)$, and the total count $\sum_v\rho_v$ is conserved. Each step composes the target, relaxation and shift operations. The target allocates the node count among the node's slots,
\begin{equation}
\label{eq:graphtarget}
\begin{gathered}
\Eq(v,r)=\rho_v\,a_r(v),\\ a(v)=\softmax\bigl[\hat n_\theta(\rho_v,\{\rho_w\}_{w\in \mathcal N(v)},\chi_{v\cdot})\bigr],
\end{gathered}
\end{equation}
where $\hat n_\theta$ is a perceptron of the node count, the counts $\rho_w$ of its neighbors $w\in\mathcal N(v)$ and its outgoing capacities $\chi_{v\cdot}$. By construction, $\Eq(v,0)+\sum_r\Eq(v,r)=\rho_v$, and $\Eq$ depends on $z$ through the counts.
The first gives conservation under relaxation, and the two together make the target a projector.
The relaxation $z\to(1-\gamma)z+\gamma\Eq$ has a learned rate $\gamma\in(0,2)$ corresponding to the relaxation rate of a residual block.
Since the graph has no prescribed continuation direction, the shift moves the population in the edge slot $r_{vw}$ from $v$ to $w$, where it occupies the reverse-edge slot $r_{wv}$. The shift is therefore a fixed involution, swapping the two slots of each edge without learned parameters. For comparison, we also train a message-passing network of the same size and training budget, without enforcing any of the three properties.

We first test whether the graph machine generates the long-time behavior of the process from short training windows.
The error $e_\rho$ is defined as the rms count error over nodes, normalized by the rms of the data and averaged over the held-out trajectories. The learners are rolled out for $\ccTlong$ steps, ten times the length of the training window.
The machine's error at the end of the rollout is $\ccErrMThree\%$, against $\ccPersThree\%$ for leaving the counts unchanged (Fig.~\ref{fig:caseC}(A)), and its total count is conserved to rounding precision. The message-passing network, without the three properties, reaches $\ccErrBThree\%$ at the same time and does not conserve the total count. The same behavior persists in a separate run with four times the training budget (Supplementary Material, section S4 and Fig.~S4).
The machine's one-step Jacobian at a uniform state has no eigenvalue of modulus above one, and it separates a kinetic cluster at $|1-\gamma|$ from a small number of slow count modes with the slowest at an eigenvalue $\lambda_1=\ccJacMachine$ (Supplementary Material, section S4).

This separation suggests a slow manifold, on which the initial kinetic allocation is forgotten.
To test it, we start two runs with different random allocations of the same counts over the edge slots and follow the norm $|\delta z|$ of their difference relative to its initial value $|\delta z_0|$.
After one step the two runs differ by the factor $\ccDecayFirst$, the learned $|1-\gamma|$, and that first step leaves an imprint $\delta\rho$ on the counts, which then decays through the slow count modes (Fig.~\ref{fig:caseC}(B) and Supplementary Material, section S4). By contrast, a perturbation of the message-passing network's hidden channels grows (Fig.~\ref{fig:caseC}(B)).
With $\gamma$ fixed at $2.0$, the machine loses the third property, and the difference between the two allocations grows. %

The emergent law can be read from the trained machine. Let $z^*$ denote the post-collision state and $J_{vw}=z^*(v,r_{vw})-z^*(w,r_{wv})$ the net flux on each edge. Because the target depends only on the counts and the shift is a reflection, the counts and these fluxes obey an exact closed law (Supplementary Material, section S4). This reduction contains one kinetic variable per edge and a memory term from the previous step and reproduces the machine to rounding for every $\gamma$ tested. The kinetic variable relaxes with eigenvalue $\gamma-1$ and is slaved to the counts on the slow count modes,
\begin{equation}
\label{eq:graphlaw}
J_{vw}\approx\kappa\bigl[\Eq(v,r_{vw})-\Eq(w,r_{wv})\bigr](\rho),\qquad \kappa=\frac{\gamma}{2-\gamma},
\end{equation}
giving a first-order conservation law for the counts alone, with the learned allocation as flux. The coefficient $\kappa$ arises from the memory term retained by the reflection and is set by the kinetic eigenvalue, in the same way that the diffusivity of the scalar machine is set by $1-\omega$. In the CTG sectors, the net edge flux is the transport variable of the graph machine, while the symmetric combination of the two edge slots and the rest variable form its ghosts, which do not enter the count dynamics. %

The error $e_{\rm law}$ of the law \eqref{eq:graphlaw} is measured along the held-out trajectories using the same measure as $e_\rho$, now applied to the difference between the counts of the law and those of the machine. With $\kappa=\ccFluxFactor$ determined by the learned $\gamma=\ccGamma$ and no additional fitting, $e_{\rm law}$ is $\ccLawRenThirty\%$ at $t=30$, whereas setting $\kappa=1$ as for a direction-preserving shift gives an error of $\ccLawLeadThirty\%$ (Fig.~\ref{fig:caseC}(C)). A machine trained with $\gamma$ fixed at $0.5$ confirms the coefficient for a relaxation rate below one (Supplementary Material, section S4).
Only the divergence of the flux is identifiable from the counts because flows around the cycles of the graph leave the counts unchanged. This cycle part of the transport sector is the graph counterpart of the weakly constrained eigenvalues of the learned collision.

To identify the role of each property, we run three ablations, changing one construction element at a time in the same network and retraining on the same data with everything else unchanged (Supplementary Material, Fig.~S4). Without conservation, the allocation need not sum to the node count, and the network must learn the count balance from the data. It learns this balance on average but not node by node. The total count changes by up to $\ccAbOneMassThree\%$ at $t=\ccTlong$, and the reduction to counts and fluxes misses the network by \ccAbOneClosure\% without a source term. A property that the data satisfy is therefore learned approximately where it is not imposed, and the count balance acquires a learned source.

With $\gamma$ fixed at $2.0$ to remove the gap, the kinetic part $K(t)=|z-\Eq(\rho)|$ does not decay, and at $\gamma=2.3$ it grows. Two allocations of the same counts grow apart (Fig.~\ref{fig:caseC}(B)), and the first-order approximation fails in this test, since a conservative law of the machine's form fitted to the rollouts misses them by $\ccAbTwoFitLaw\%$. Here the third property fails on the graph through the loss of kinetic damping at $\gamma=2$ rather than through mixing at $\Pi'=1$ on the lattice.

Without the projector, the target can depend on the allocation. Training nevertheless keeps it within $\ccAbThreeDefect\%$ of a projector, while the exact reduction breaks by \ccAbThreeClosure\%. This ablation has the smallest effect since the learned target remains nearly independent of the kinetic allocation in the process with no kinetic variables.

The ablations link conservation to the exact count balance, the gap to the slow manifold and the first-order law in the counts, and the projector to the exact count-flux reduction. The properties, which the lattice gives LB by construction, produce the same behavior when imposed directly on a graph.

\section*{Discussion}

We have placed the lattice Boltzmann method in the class of iterated relaxation dynamics to which residual networks and transformers belong and identified three properties of its update rules that allow the iteration to generate a closed macroscopic law. The three properties and the kinetic spectrum determine whether and where the law emerges, while the particular target determines its form. A D1Q3 machine with a random constrained target generated the predicted nonlinear transport law with errors of about $10^{-5}$ in its dispersion relation and below $10^{-4}$ in its nonlinear profiles. Its fronts traveled at the Rankine--Hugoniot speed of a randomly drawn flux. At the boundaries tested on the scalar lattice and the graph, the mechanism failed whenever the composed step ceased to damp a non-conserved mode, and on the lattice the macroscopic law gave no warning of such failure. In the CTG sectors of the state, the conserved sector is what the machine conserves. The transport sector is what it transports along, which sets the dissipative coefficients of the law, and macroscopic data identify only the part of it that enters the law. The ghost sector is what the machine must damp (scale separation), even though its presence is necessary to secure the correct symmetries of the higher-dimensional phase-space representation. On the lattice the checkerboard instability arose through this sector, by its coupling to the conserved sector at $k=\pi$, whereas on the graph the loss of damping at $\gamma=2$ affected the transport variable as well.

LB then differs from the networks it resembles in how its operators are specified. The shift is fixed by the lattice and the target is constrained by conservation and symmetry, so the operators that a network learns through training are prescribed in LB by physics. Within this structure, new physics enters through changes to the equilibrium, forcing or relaxation in the collision step while the propagation step remains unchanged. Whether these changes come from a physical model or from data, the appropriate conservation and balance laws, the target's projector property and the gap of the composed step need to be preserved. The projection in \eqref{eq:learned} and the softmax target in \eqref{eq:graphtarget} illustrate how conservation and the projector property can be imposed by construction, whereas the gap still needs to be verified for the composed step. Macroscopic data can constrain the combinations of kinetic rates that enter the emergent coefficients at the sampled wavenumbers, while kinetic directions that barely affect the observed fields remain weakly constrained, as in our learned collision. This is an identifiability limit that a learned collision operator \cite{Corbetta2023,Ortali2024,Bedrunka2024} can encounter when it is trained on fields rather than on populations. Differentiable implementations such as XLB \cite{Ataei2024} and Lettuce \cite{Bedrunka2021}, which allow gradients to pass through every collision and shift as they did here, let an LB step act as a layer in the same software used to train deep networks.

Our machines are small. The random target is scalar and one-dimensional, and its nonlinear tests were run at one relaxation rate. The learned collision was identified in a one-dimensional geometry where the law has two coefficients. The graph machine was trained on one graph using windows mostly near the process's fixed point. The mechanism concerns the slow manifold and the emergent closed law but does not establish the stability of a full scheme with boundaries. The correspondence with networks also remains structural. Within those limits, the lattice Boltzmann machine is generative in a sense that can be stated formally and numerically tested. A graph given the three properties by hand shows what the lattice gives LB by construction. It is hoped and expected that the formal connection between the lattice Boltzmann machine and transformer architectures discussed in this paper will help promote further cross-fertilization between these two important areas at the forefront of modern statistical physics and computational science.

\paragraph{Data, Materials, and Software Availability.} The code and data that support the findings of this study are available from the corresponding author upon reasonable request.

\paragraph{Acknowledgments.} The research of H.-D.Y. is financed by the Swedish Transport Administration in the project ``GEneric Multidisciplinary optimization for sail INstallation on wInd-assisted ships'' (GEMINI, Grant No.~TRV~2023/32107).

\paragraph{Author contributions.} H.-D.Y. and S.S. designed research; H.-D.Y. performed research; H.-D.Y. and S.S. wrote the paper.

\paragraph{Competing interests.} The authors declare no competing interest.

\clearpage
\setcounter{figure}{0}
\setcounter{equation}{0}
\setcounter{table}{0}
\renewcommand{\thefigure}{S\arabic{figure}}
\renewcommand{\theequation}{S\arabic{equation}}
\renewcommand{\thetable}{S\arabic{table}}
\renewcommand{\theHfigure}{S\arabic{figure}}
\renewcommand{\theHequation}{S\arabic{equation}}
\renewcommand{\theHtable}{S\arabic{table}}

\begin{center}
{\Large\bfseries Supplementary Material}\\[6pt]
{\large The lattice Boltzmann machine}\\[4pt]
Hua-Dong Yao and Sauro Succi
\end{center}

\bigskip

This Supplementary Material gives the derivations supporting the generative mechanism in the main text, together with the set-up, protocols, additional results and robustness checks for the three machines. Section S1 derives the results for the class and the scalar lattice. Section S2 covers the scalar machine studied in \textit{Emergence as generative mechanism} and \textit{Where the mechanism stops} (Figs.~S1 and S2). Section S3 gives further details of \textit{Learning collision through the machine} (Fig.~S3), and section S4 of \textit{Three properties tested on a graph}. Equations from the main text retain their numbers. Equations in this Supplementary Material are numbered S1 onward.

\subsection*{S1. Derivations for the class}

The projector identities below hold for every member of the class with a projector target. The subsequent expansion assumes a lattice shift, a smooth target, a single constant relaxation rate and long-wavelength ordering. We then derive the checkerboard channel and the three sectors for the D1Q3 machine, where both can be expressed in closed form. The section closes with the comparison to attention referred to in \textit{The common class of iterated relaxation dynamics} in the main text.

\paragraph{The projector and the collision Jacobian.}
Let $\Eq=\Eqf\circ\Mom$ and $\Mom\Eqf(m)=m$, and let $D$ denote the derivative with respect to the displayed argument. Differentiating the moment identity gives $\Mom\,D\Eqf(m)=\Id_{\rm c}$, where $\Id_{\rm c}$ is the identity on the conserved moments. The equilibrium Jacobian therefore satisfies
\begin{equation}
\Proj(f)=D\Eqf(\Mom f)\,\Mom,\qquad \Mom\Proj=\Mom,\qquad \Proj^2=D\Eqf\,(\Mom D\Eqf)\,\Mom=\Proj.
\end{equation}
These identities hold at every admissible state. For a multiple-relaxation-time collision $\Col(f)=f+\Rel[\Eq(f)-f]$, let $\Rel=\mathsf B^{-1}\mathrm{diag}(\omega_k)\mathsf B$, with the conserved moments among the rows of $\mathsf B$. The same calculation gives the collision Jacobian
\begin{equation}
\Col'=\Proj+(\Id-\Rel)(\Id-\Proj).
\end{equation}
Its kinetic eigenvalues are $1-\omega_k$.

\paragraph{The expansion for the class.}
The flux and second-moment tensor of the target are vectors in moment space. With $\ell$ denoting the moment index, they are defined by
\begin{equation}
\label{eq:fluxes}
\Phi^{\beta}_\ell(m)=\sum_i\Mom_{\ell i}\,c_{i\beta}\,\Eqf_i(m),\qquad
\Psi^{\beta\beta'}_\ell(m)=\sum_i\Mom_{\ell i}\,c_{i\beta}c_{i\beta'}\,\Eqf_i(m).
\end{equation}
A Taylor expansion of the shifted populations in the update, \eqmain{1}, gives
\begin{equation}
\mathcal D_if_i+\tfrac12\mathcal D_i^2f_i+\dots=-\omega(f_i-\Eq_i),\qquad \mathcal D_i=\partial_t+\vc_i\cdot\nabla.
\end{equation}
We insert the expansion $f=\Eqf(m)+\epsilon f^{(1)}+\epsilon^2f^{(2)}$, with the ordering $\partial_t=\epsilon\partial_{t_1}+\epsilon^2\partial_{t_2}$ and $\nabla=\epsilon\nabla_1$. The corrections satisfy $\Mom f^{(1)}=\Mom f^{(2)}=0$. At order $\epsilon$, the update gives
\begin{equation}
\mathcal D_{i1}\Eqf_i=-\omega f^{(1)}_i,\qquad \mathcal D_{i1}=\partial_{t_1}+\vc_i\cdot\nabla_1.
\end{equation}
Thus, $f^{(1)}$ lies in $\ker\Mom$ and is determined by the relaxation for $\omega\neq0$. Substituting this first-order result at order $\epsilon^2$ gives
\begin{equation}
\partial_{t_2}\Eqf_i+\Bigl(\tfrac12-\tfrac1\omega\Bigr)\mathcal D_{i1}^2\Eqf_i=-\omega f^{(2)}_i.
\end{equation}
Applying $\Mom$ at first order gives $\partial_{t_1}m+\partial_\beta\Phi^\beta=0$. At second order, it gives
\begin{equation}
\partial_{t_2}m_\ell=\Bigl(\tfrac1\omega-\tfrac12\Bigr)\sum_i \Mom_{\ell i}\mathcal D_{i1}^2\Eqf_i,\qquad
\sum_i \Mom_{\ell i}\mathcal D_{i1}^2\Eqf_i=\partial_{t_1}^2m_\ell+2\partial_{t_1}\partial_\beta\Phi^\beta_\ell+\partial_\beta\partial_{\beta'}\Psi^{\beta\beta'}_\ell.
\end{equation}
Using $\partial_{t_1}m=-\partial_\beta\Phi^\beta$ in the first term reduces the sum to $\partial_\beta[\partial_{t_1}\Phi^\beta+\partial_{\beta'}\Psi^{\beta\beta'}]$, which yields \eqmain{6}. For the scalar D1Q3 machine, $\Phi=F$, $\Psi=\Pi$ and $\partial_{t_1}F=F'\partial_{t_1}\rho=-F'\partial_xF$. The expression in brackets then becomes
\begin{equation}
\partial_x\Pi-F'\partial_xF=(\Pi'-F'^2)\,\partial_x\rho.
\end{equation}
This gives \eqmain{8}. For the linear equilibrium $\Eqf_i=w_i(1+c_iu/c_s^2)\rho$ with $c_s^2=1/3$, the result is $D=(1/\omega-1/2)(1/3-u^2)$, the known finite-velocity diffusivity of the advection--diffusion scheme \cite{SI-Huang2014}.

\paragraph{The growth factor and the checkerboard eigenvalue.}
For a member of the class with lattice velocities $\vc_i$, the growth factor of one step linearized about a uniform state at wavenumber $\mathbf k$ is $A(\mathbf k)=\mathrm{diag}(e^{-i\mathbf k\cdot\vc_i})\,\Col'$. In the D1Q3 machine, let $\xi_i=\Eqf_i'(\rho_0)$. The collision matrix is then $(1-\omega)\Id+\omega\,\xi\vone^\top$, and the growth factor at wavenumber $k$ is
\begin{equation}
A(k)=\mathrm{diag}(e^{-ikc_i})\bigl[(1-\omega)\Id+\omega\,\xi\vone^\top\bigr].
\end{equation}
The eigenvalue continuously connected to $1$ at $k=0$ is $\exp[-iF'(\rho_0)k-D(\rho_0)k^2+O(k^3)]$. We use this relation to check \eqmain{8} against the exact lattice dispersion in Fig.~\ref{fig:S1}(C) and (D). Figure~\ref{fig:S2}(A) shows the three branches of $A(k)$ for the first random machine at $\rho_0=0.4$ and $\omega=0.8$. The hydrodynamic branch leaves $1$, while the kinetic branches remain near $1-\omega$.

If all components of $\xi$ are positive, the growth factor amplifies no mode at a uniform state. The projector $P=\xi\vone^\top$ is then orthogonal in the inner product weighted by $1/\xi_i$, in which the shift is unitary. With $Q=\Id-P$,
\[
\|A(k)v\|_W^2=\|Pv\|_W^2+(1-\omega)^2\|Qv\|_W^2\le\|v\|_W^2
\]
for every wavenumber and $0<\omega<2$, where $\|v\|_W^2=\sum_i|v_i|^2/\xi_i$. A mode of unit modulus would need an eigenvector proportional to $\xi$ and equal streaming phases for the three velocities, which on D1Q3 happens only at $k=0$. The shape constraints of section S2 make $\xi_\pm=(\Pi'\pm F')/2$ positive, and the two accepted targets keep $\xi_0=1-\Pi'$ positive over the whole density range. The rejected target loses this property at $\Pi'=1$, where the checkerboard eigenvalue derived below reaches $-1$.

At $k=\pi$ on D1Q3, the growth factor is $A(\pi)=\Sigma[(1-\omega)\Id+\omega\,\xi\vone^\top]$, with $\Sigma=\mathrm{diag}(-1,1,-1)$. The eigenvalues affected by the rank-one update of $(1-\omega)\Sigma$ satisfy the secular equation
\begin{equation}
1+\omega\sum_i\frac{\Sigma_{ii}\xi_i}{(1-\omega)\Sigma_{ii}-\lambda}=0.
\end{equation}
Substituting $\xi_{\pm}=(\Pi'\pm F')/2$ and $\xi_0=1-\Pi'$ gives the quadratic equation
\begin{equation}
\lambda^2-\omega(1-2\Pi')\lambda-(1-\omega)=0.
\end{equation}
Its roots are
\begin{equation}
\label{eq:lampi}
\lambda_\pm=\tfrac12\Bigl[\omega(1-2\Pi')\pm\sqrt{\omega^2(1-2\Pi')^2+4(1-\omega)}\Bigr].
\end{equation}
Here $\Pi'$ is evaluated at the uniform density. The third eigenvalue is $\omega-1$, on the vector $(1,0,-1)$ that $\Sigma$ reverses. Setting $\lambda=-1$ gives $\omega(2-2\Pi')=0$, so $|\lambda_-|=1$ if and only if $\Pi'=1$. For the slow-transport targets of the random machine, the checkerboard mode is the first to lose contraction. Targets with faster transport can lose stability at intermediate wavenumbers for rates above one, which this closed form does not cover.

\paragraph{The three sectors with separate rates.}
For the D1Q3 machine, take moments with the rows $(1,1,1)$, $(-1,0,1)$ and $(1,-2,1)$, corresponding to the conserved, transport and ghost sectors. Let the collision relax the transport and ghost moments at rates $\omega_T$ and $\omega_G$. The target moments in this basis are $(\rho,F,3\Pi-2\rho)$, so one collision acts on the linearized moments as
\begin{equation}
m_C\to m_C,\qquad m_T\to(1-\omega_T)\,m_T+\omega_T F'\,m_C,\qquad m_G\to(1-\omega_G)\,m_G+\omega_G(3\Pi'-2)\,m_C.
\end{equation}
The two kinetic sectors are coupled to the conserved sector but not to each other. In the expansion above, $f^{(1)}$ enters the law only through the fluxes of the conserved moments, which depend on its transport component. The coefficient of the law is therefore $(1/\omega_T-1/2)(\Pi'-F'^2)$ for any ghost rate. A numerical check with the exact growth factor gives this value to better than $10^{-7}$ over fifty random targets and rate pairs. The ghost rate first changes the slow eigenvalue at third order in $k$ when $F'\neq0$ and at fourth order when $F'=0$.

At $k=\pi$, the shift $\Sigma=\mathrm{diag}(-1,1,-1)$ reverses the transport moment and mixes the conserved and ghost moments according to
\begin{equation}
\Sigma(1,1,1)=-\tfrac13(1,1,1)-\tfrac23(1,-2,1),\qquad \Sigma(1,-2,1)=-\tfrac43(1,1,1)+\tfrac13(1,-2,1).
\end{equation}
The step at $k=\pi$ is therefore block triangular. The transport moment is driven by the conserved moment through $F'$ but does not feed back into it. The eigenvalues are $\omega_T-1$, on the exact eigenvector $(1,0,-1)$, and the roots of $\lambda^2-\omega_G(1-2\Pi')\lambda-(1-\omega_G)=0$. These roots are given by \eqref{eq:lampi} with $\omega$ replaced by $\omega_G$, so the threshold $\Pi'=1$ holds for every transport and ghost rate in $(0,2)$. The transport mode, reversed by the shift, drives neither of the other two sectors, and $\lambda_\pm$ depends on the ghost rate alone. The eigenvectors of $\lambda_\pm$ carry a transport component proportional to $F'$, slaved to the conserved and ghost components. The checkerboard instability therefore arises from a loop between the conserved and ghost sectors, in which the transport sector takes no active part.

For the graph machine, the net edge flux $J$ is the transport variable because it alone enters the count change. The symmetric edge combination and the rest variable are its ghosts. They relax at $1-\gamma$ and do not enter the count change (section S4).

\paragraph{The attention operation and the pseudopotential interaction.}
The equilibrium of the LB machine mixes populations over the velocity index at a single site. Attention \cite{SI-Vaswani2017} mixes token representations across positions as $\sum_\nu\alpha_{\mu\nu}(z)\,W_Vz_\nu$. The coefficients $\alpha_{\mu\nu}(z)$ are formed from learned projections of the state, and $W_V$ is the value projection. A plain layer instead mixes with a learned dense kernel $W$. The nearest LB counterpart of attention is therefore the interaction between sites. In the pseudopotential interaction \cite{SI-Shan1993}, a density-dependent force between neighboring sites enters the collision through a shift of the equilibrium current. This gives a state-dependent coupling across sites, restricted to one shell and with one parameter in place of the learned projections.

\subsection*{S2. The scalar machine}

The scalar machine in \textit{Emergence as generative mechanism} in the main text is a D1Q3 machine with a random constrained target. In \textit{Where the mechanism stops}, the same lattice is given a target that fails the stability test and is driven past its stability limit. This section gives the set-up and protocols, additional results for the two targets and the relaxation-rate scan, and tests with relaxed conservation and with a state-dependent rate.

\paragraph{Set-up and protocols.}
The machine uses D1Q3 on a periodic lattice at $\omega=\cznOmega$, over the density range $[0.2,1]$. The target moments $F$ and $\Pi$ are cubic polynomials with coefficients drawn from a seeded normal generator. They are rescaled so that $0.05\rho\le\Pi\le0.95\rho$, $|F|\le0.9\Pi$, $|F'|\le0.1$ and $\Pi'-F'^2\ge0.2$ throughout the range. A draw is retained only if the growth factor of the linearized step has no eigenvalue of modulus above one at any of 17 uniform densities and 64 wavenumbers. The two targets in the main text are the first admissible draws from their seeds, obtained after \cznTrialsOne{} and \cznTrialsTwo{} trials.

The dispersion runs start from a sinusoidal perturbation of relative amplitude $10^{-4}$ on 1024 sites, using modes 1, 2, 4, 8 and 16. The decay rate and phase speed are obtained by fitting the last three quarters of each run. The relaxation-rate scan uses 256 sites, mode 4, $\rho_0=0.6$ and $\omega$ from $0.8$ to $1.98$.

The nonlinear-profile runs use 1024 sites and 12000 steps, starting from $\rho=0.6+0.336\sin(2\pi x/N)\exp[-(x-N/2)^2/2(0.22N)^2]$. This profile spans densities from $0.38$ to $0.82$. We compare the machine with a conservative finite-volume solution of \eqmain{8} on a grid refined by a factor of two, using central fluxes and forward Euler time stepping. Let $\Delta x$ denote the grid spacing, and let $D_{\max}$ and $\max|F'|$ be the maxima of the diffusivity and the magnitude of the flux derivative over the density range $[0.2,1]$. The time step is chosen with a safety factor of 0.4,
\begin{equation*}
\Delta t=0.4\min\!\left(\frac{\Delta x^2}{2D_{\max}},\frac{\Delta x}{\max|F'|}\right).
\end{equation*}
The step is then adjusted slightly to reach the final time exactly. Diffusion controls the step in both reported runs. The advective Courant numbers are approximately 0.0120 for the first machine and 0.00621 for the second. Increasing the refinement factor to four changes the solution by less than \cznFvCheck{} at $t=3000$.

The front runs start from smoothed steps between two densities chosen by the Oleinik chord condition. Front speeds are measured from the half-height crossing over records 2 to 10 of 14.

The conservation control adds $\Delta\rho$ to the rest population of the target, giving $\sum_i\Eqf_i=(1+\Delta)\rho$ while leaving $F$ and $\Pi$ unchanged. It uses $\Delta=-10^{-4}$ and runs for 2000 steps.

The instability runs use a draw of seed 23 that meets the shape constraints but fails the stability test. The second machine in the main text is a later, stable draw from the same seed. We use the same $\omega$ on 512 sites for 400 steps. The initial profiles have plateaus between $x=0.25N$ and $0.75N$, transitions of width 16, and a seeded checkerboard kinetic perturbation of amplitude $10^{-9}$ along $(1,-2,1)/\sqrt6$. The mode amplitude is measured with a Hann window over the central 30\% and fitted over the range $10^{-8}$ to $10^{-3}$.

The state-dependent-rate runs use 256 sites, mode 2, $\rho_0=\cbRhoZero$ and a uniform kick along the same direction, with $\alpha\zeta=\omega(f_0)-\omega_0$ at $\rho_0$. The decay rate is measured over steps 500 to 6000.

\begin{figure}[!htb]%
\centering
\includegraphics[width=\linewidth]{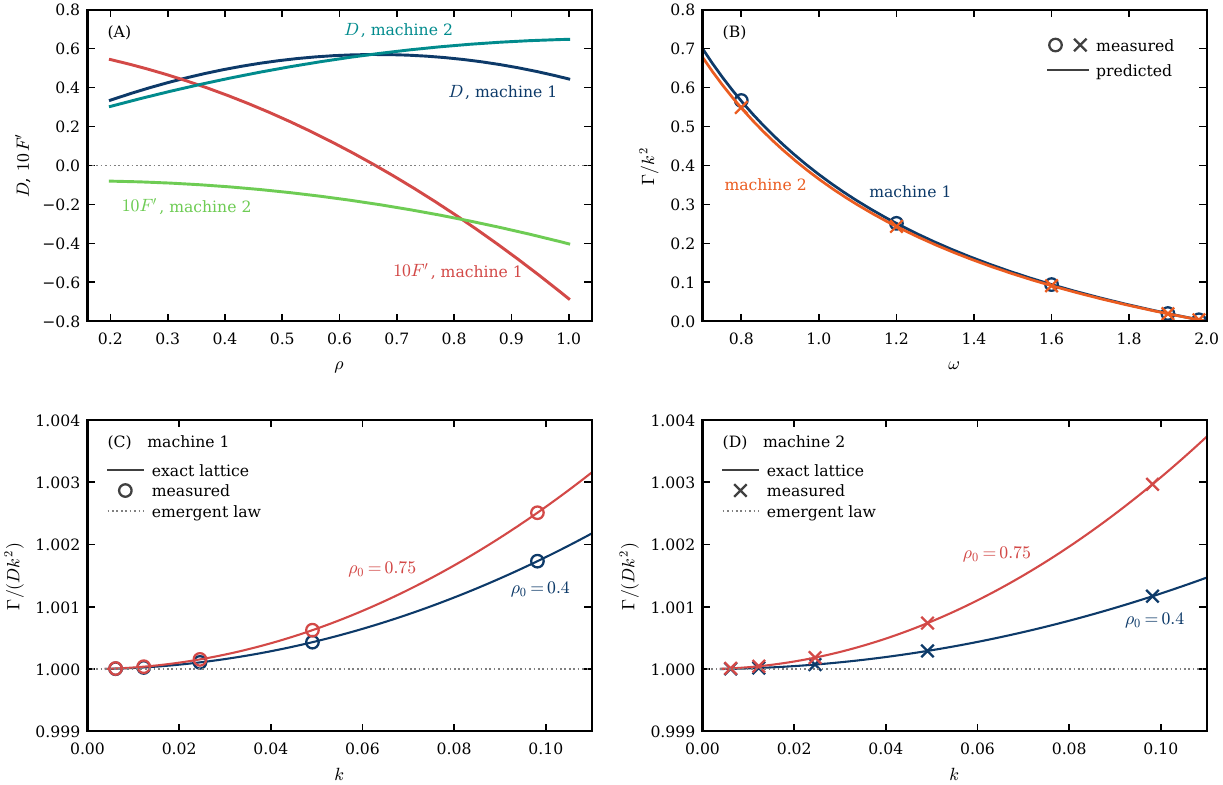}
\caption{The two random targets and the linear response of the scalar machine. (A) Diffusivity $D(\rho)$ and ten times the flux derivative $F'(\rho)$ for the two targets at $\omega=\cznOmega$. (B) Decay rate of the fourth mode on 256 sites, divided by $k^2$, against $\omega$ for the first machine (circles) and the second machine (crosses) at $\rho_0=0.6$, with the predictions $(1/\omega-1/2)[\Pi'-F'^2]$ (lines). (C and D) Measured decay rate $\Gamma$ of a sinusoidal perturbation about $\rho_0=0.4$ and $0.75$, divided by the predicted $D(\rho_0)k^2$, against wavenumber $k$ for the first machine (C, circles) and the second machine (D, crosses). Solid curves show the exact lattice growth factor, and the dotted line shows the emergent law.}
\label{fig:S1}
\end{figure}

\paragraph{The two random targets and the relaxation-rate scan.}
Figure~\ref{fig:S1}(A) shows the flux derivatives and diffusivities of the two random targets used in the main text. Figure~\ref{fig:S1}(B) shows the measured decay rate of the fourth mode of both machines on 256 sites, divided by $k^2$, over the relaxation-rate range $\omega=0.8$ to $1.98$. The prediction is $(1/\omega-1/2)[\Pi'-F'^2]$ at $\rho_0=0.6$. The measured coefficients follow the predicted dependence on the relaxation rate through the factor $1/\omega-1/2$ throughout the range, with deviations of \cznOmegaDevLast{} and \cznOmegaDevLastTwo{} for the first and second machine at $\omega=1.98$. At this rate, the kinetic modes decay in about fifty steps and the diffusivity is a factor $\cznDRatio$ below its value at $\omega=0.8$. The largest departures over the scan, \cznOmegaDev{} and \cznOmegaDevTwo{} at $\omega=\cznOmegaDevAt$, are the finite-wavenumber correction for the mode used.

Figures~\ref{fig:S1}(C) and (D) compare the measured decay rate of a sinusoidal perturbation with the predicted $D(\rho_0)k^2$ for both machines at two base densities, together with the exact lattice growth factor $A(k)$. The dispersion runs cover wavenumbers from $\cznKmin$ to $\cznKmax$. At the smallest wavenumber, the phase speed agrees with $F'(\rho_0)$ to \cznSpeedDevSmall{} in relative terms. At every wavenumber, the measured decay rate agrees with the hydrodynamic eigenvalue of $A(k)$ to five digits. The departure from the emergent law at larger wavenumber, \cznDevLarge{} at $k=\cznKmax$, is therefore the finite-wavenumber correction neglected by the expansion. The spectrum of the composed step determines the linear response of the machine.

\paragraph{The second machine and the fronts.}
In the nonlinear run of the main text, the rms difference between the solution of the law linearized about the mean density and the first machine's density profile is \cznRmsLinOne{} of the predicted profile's standard deviation at the final time. For the second machine, the rms differences from the solutions of its predicted and linearized laws are \cznRmsTwo{} and \cznRmsLinTwo{}, respectively. Both are normalized by the standard deviation of the second machine's predicted profile at the same final time. The two refinements of the finite-volume reference differ by less than \cznFvCheck{} in density at $t=3000$.

The front profile is set by the balance between steepening by the nonlinear flux and spreading by diffusion \cite{SI-Nicolis1977}. The fronts start from smoothed steps between densities chosen by the Oleinik chord condition, $\rho=\cznRhoLOne$ and $\cznRhoROne$ for the first machine and $\cznRhoLTwo$ and $\cznRhoRTwo$ for the second. The measured front speed is $\cznSmeasOne$ for the first machine, compared with the Rankine--Hugoniot prediction $\cznSrhOne$ from its random flux. For the second machine, the corresponding values are $\cznSmeasTwo$ and $\cznSrhTwo$.

\paragraph{Conservation relaxed.}
Adding $\Delta\rho$ to the rest population of the first machine's target gives $\sum_i\Eqf_i=(1+\Delta)\rho$, with $\Delta=-10^{-4}$. Over 2000 steps, the total mass changes by a factor of $\cznConsMeas$, compared with the predicted $(1+\omega\Delta)^{2000}=\cznConsPred$. This change follows from the source term acquired by the expansion when conservation is relaxed. Relaxing the first property therefore changes the law rather than removing it.

\begin{figure}[!htb]%
\centering
\includegraphics[width=\linewidth]{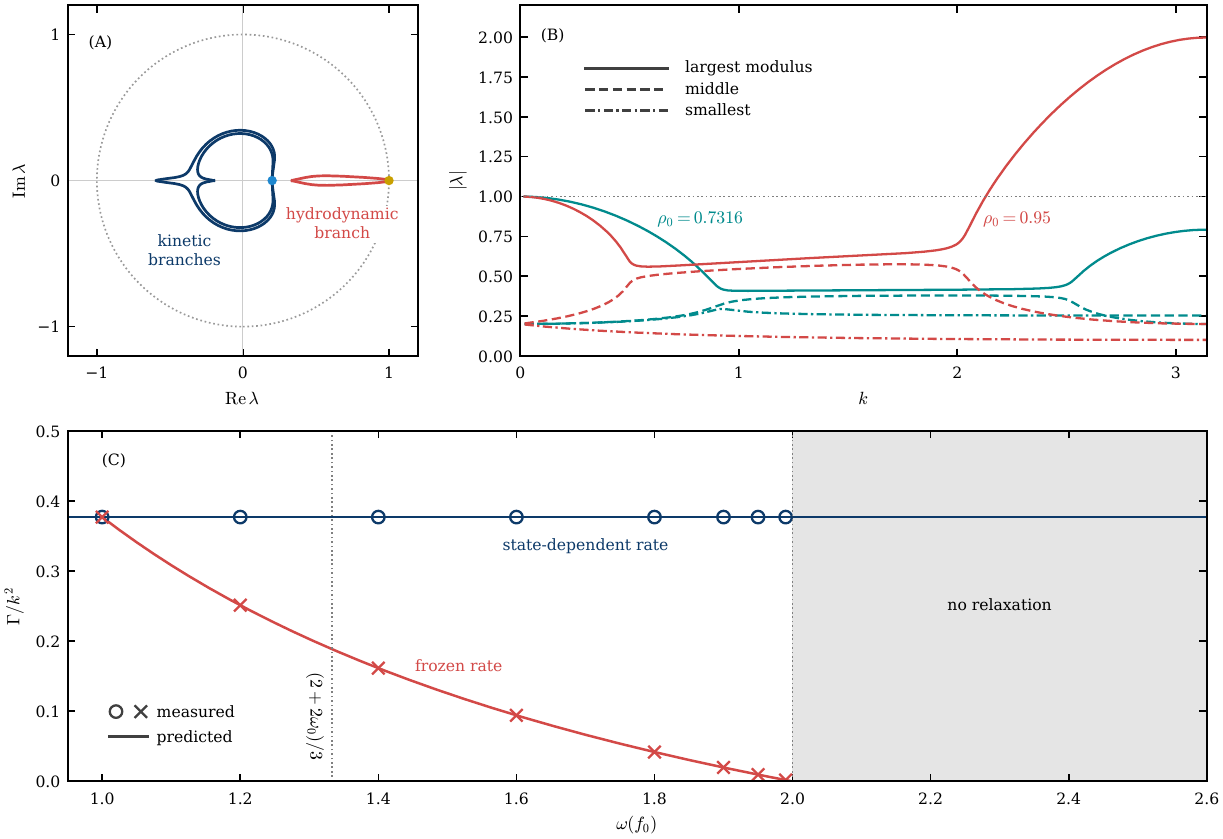}
\caption{Spectra of the scalar machine and the state-dependent rate. (A) Spectrum of the composed step for the first random machine in the complex plane at $\rho_0=0.4$ and $\omega=0.8$. Circles mark the conserved eigenvalue $1$ and kinetic eigenvalue $1-\omega$ of the collision. The curves show the three branches of $A(k)$ for $k$ from $0$ to $\pi$. The hydrodynamic branch leaves $1$ (red), while the kinetic branches remain near $1-\omega$ (blue). (B) Moduli of the three eigenvalues of $A(k)$, ordered from largest to smallest at each $k$, against $k$ at the two plateau densities of the instability runs. (C) Decay rate of a density wave, divided by $k^2$, against the initial instantaneous rate $\omega(f_0)$ evaluated at $\rho_0$, after a uniform kinetic kick. Circles denote the state-dependent rate and crosses the rate frozen at its initial value. Lines show the predictions $D(\rho_0;\omega_0)$ and $D(\rho_0;\omega(f_0))$. Both machines diverge beyond $\omega(f_0)=2$ (shaded). The dotted line marks $(2+2\omega_0)/3$.}
\label{fig:S2}
\end{figure}

\paragraph{The instability runs.}
Figure~\ref{fig:S2}(B) shows the moduli of the three eigenvalues of $A(k)$ at the two plateau densities used in the instability runs, ordered from largest to smallest at each $k$. At $\rho_0=\cbRhoHiStable$, every modulus remains below one for $k>0$, with $|\lambda_-|=\cbLamStable$. At $\rho_0=\cbRhoOne$, the largest modulus reaches $|\lambda_-|$ at $k=\pi$, while the predicted law remains regular with $D=\cbDRhoOne$. At both plateau densities, $\lambda_-$ is the end of the hydrodynamic branch at $k=\pi$.

Below $\rho_c$, the checkerboard amplitude falls by \cbAmpDropTwenty{} in twenty steps, below $|\lambda_-|^{20}=9\times10^{-3}$. This faster initial decay occurs because the seeded ghost direction lies mostly on the faster coupled mode $\lambda_+$ and only partly on $\lambda_-$. The transport mode is orthogonal to the measured direction. The amplitude then settles at a floor of \cbFloorEnd{}, caused by the smooth kinetic part of the plateau edges leaking into the windowed measurement. The density follows the solution of \eqmain{8} to \cbRmsStable{} over \cbTStable{} steps. Above $\rho_c$, the run diverges after \cbStepsUnstable{} steps.

\paragraph{The state-dependent rate.}
When the relaxation rate depends on the kinetic part, the condition for relaxation depends on the instantaneous rate. Entropic LB provides an example, with the relaxation rescaled at every site and step so that the discrete entropy does not decrease \cite{SI-Ansumali2002}. For a spatially uniform kinetic part, which the shift leaves unchanged, a kick relaxes only when the instantaneous rate is below two. The emergent coefficient is then set by the rate to which the machine returns on the manifold. We test this distinction by applying a uniform kinetic kick to the first machine.

Let $\omega(f)=\omega_0+\alpha\zeta$, where $\zeta=|f-\Eq(f)|^2/\rho^2$, $\alpha\ge0$ and $0<\omega_0<2$. Write the kinetic part as $g=f-\Eq(f)$. Since $\Eq$ depends only on $\Mom f$ and collision preserves $\Mom f$, the collision acts on $g$ exactly as
\begin{equation}
\Col(f)-\Eq(f)=[1-\omega(f)]\,g,\qquad \zeta'=[1-\omega_0-\alpha\zeta]^2\,\zeta.
\end{equation}
Thus, $|g|$ contracts if and only if $\omega(f)<2$, and $\zeta$ converges to zero for every $\alpha\zeta<2-\omega_0$. The derivative of the map along $g$ is $1-\omega(f)-2\alpha\zeta$, which reaches $-1$ at $\omega(f)=(2+2\omega_0)/3$. A derivative below $-1$ separates neighboring nonequilibrium states for a few steps while both relax, without changing the basin.

The basin of the slow manifold under collision is therefore bounded by the instantaneous rate reaching two. Inside it, the rate falls back to $\omega_0$ as the kinetic part relaxes. A spatially uniform kinetic kick on a uniform state is unchanged by the shift and should relax for every initial rate below two. When superposed on a density wave, the kick should relax and the wave should then decay with the diffusivity set by the rate on the manifold, $D(\rho_0;\omega_0)$. If the rate is instead frozen at its initial value, the same wave should decay with $D(\rho_0;\omega(f_0))$, where $f_0$ is the initial state. This machine should lose kinetic damping at the same threshold, since its kinetic eigenvalue $1-\omega$ then has modulus one or more.

Figure~\ref{fig:S2}(C) confirms these predictions for the first machine at $\rho_0=\cbRhoZero$, $\omega_0=\cbOmegaZero$ and $\alpha=\cbAlpha$. The kick is superposed on a density wave of relative amplitude $10^{-3}$, and the initial rate $\omega(f_0)$ is evaluated at $\rho_0$. The state-dependent machine gives $\Gamma/k^2=\cbDZero$ for every initial rate from $\cbOmegaZero$ to $\cbLastStable$, with a spread of \cbStateSpread. The frozen-rate machine follows $(1/\omega-1/2)[\Pi'-F'^2]$ down to $\cbFrozenLast$ at $\cbLastStable$. Both diverge from $\cbFirstDiv$. Neither shows a change at $(2+2\omega_0)/3$, where the derivative of the collision map along $g$ reaches $-1$. For a state-dependent rate and a uniform kinetic part, the instantaneous rate therefore sets the boundary of relaxation, while the rate on the manifold sets the slow manifold and its emergent coefficient.

\subsection*{S3. The learned collision}

The learned collision in \textit{Learning collision through the machine} in the main text is a D2Q9 collision correction trained through the machine on density and current fields. This section gives the set-up and protocols, the control and conservation checks, and the interpretation of the effective longitudinal eigenvalue. It also examines how far the identification extends in wavenumber.

\paragraph{Set-up and protocols.}
The lattice is D2Q9 with $\caNX$ sites and fields uniform in $y$. The reference is the standard D2Q9 multiple-relaxation-time scheme with the BGK equilibrium. Its moment basis is $(\rho,J_x,J_y,e,\varepsilon,q_x,q_y,p_{xx},p_{xy})$, comprising density, current, energy, squared energy, energy flux and deviatoric stress, with rates $(s_e,s_\varepsilon,s_q,s_p)=(1.2,1.4,1.1,1.6)$. The network has architecture $9\to32\to32\to9$ with $\tanh$ activations. Its input is $f_i/\rho-w_i$, where $w_i$ are the lattice weights, and its output is scaled by $\rho$.

The training set contains 64 initial fields per training wavenumber $k_{\rm tr}$, with amplitudes drawn uniformly from $[0.005,0.02]$. The populations are initialized at equilibrium and evolved over windows of $\caTwin$ steps. Training uses Adam with a learning rate of $2\times10^{-3}$, batches of 8 and about 1700 steps per training wavenumber. The loss is differentiated through every collision and shift. Training is run in checkpointed parts, with the optimizer state carried across checkpoints.

The real parts of the eigenvalues are averaged over twelve equilibrium probe states with $\rho\in[0.8,1.2]$ and $|u_x|,|u_y|\le0.05$. Damping rates are obtained from the growth factor at $\rho=1$, with the shear branch identified by its transverse-current share and the acoustic pair by its density share. The control network is trained on one-step collision targets at 4096 random admissible states.

\paragraph{Control and conservation.}
The control network, trained on the collision itself rather than through the machine, recovers the six reference kinetic eigenvalues as $\caCtrlEig$. In the machines trained on density and current fields, the projection limits the change in conserved moments to at most \caCons{} per collision. The shift has no parameter and remains unchanged. The loss is differentiated through every collision and shift. The shift has an exact adjoint, the collision is local, and the moment projector preserves the conservation laws during training.

\paragraph{The effective longitudinal eigenvalue.}
The effective longitudinal eigenvalue varies with $k_{\rm tr}$ because finite-wavenumber data weight the energy and normal-stress channels differently. Their reference relaxation rates are $s_e$ and $s_p$, with collision eigenvalues $1-s_e=-0.2$ and $1-s_p=-0.6$, respectively. Neither collision eigenvalue is recovered on its own.

\begin{figure}[!htb]%
\centering
\includegraphics[width=\linewidth]{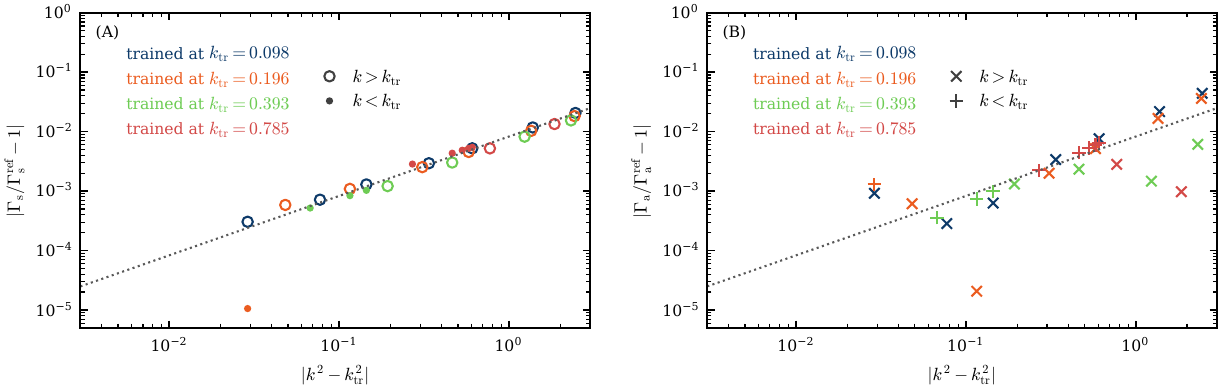}
\caption{Reach of the learned collision in wavenumber. (A) Relative error in the learned shear damping and (B) in the learned acoustic damping against $|k^2-k_{\rm tr}^2|$, the difference between the squared evaluation and training wavenumbers, for four training wavenumbers $k_{\rm tr}$. Circles and crosses mark $k>k_{\rm tr}$, while dots and pluses mark $k<k_{\rm tr}$. The dotted line has slope one and the shear-branch prefactor \caCollapseShear{}.}
\label{fig:S3}
\end{figure}

\paragraph{Reach in wavenumber.}
Figure~\ref{fig:S3} compares the shear and acoustic damping rates, $\Gamma_s$ and $\Gamma_a$, of the learned linearized step with those of the reference. It shows the relative errors for four training wavenumbers $k_{\rm tr}$. The errors are plotted against $|k^2-k_{\rm tr}^2|$, the difference between the squared evaluation and training wavenumbers. This variable follows from the dispersion relation. The damping rate of a branch has the expansion $\nu k^2+c_4k^4+\dots$, where $\nu$ is the viscosity of that branch. Training at $k_{\rm tr}$ fixes the value of the series there. When the residual at $k_{\rm tr}$ and the terms beyond $k^4$ are small, the learned and reference branches separate through the $k^4$ term, and the relative error grows in proportion to $|k^2-k_{\rm tr}^2|$.

For the shear branch, the four machines fall close to one line of slope one, with the common prefactor \caCollapseShear{} given by the median over the four. The learned shear branches therefore have nearly the same $k^4$ coefficient at every training wavenumber, set by the kinetic eigenvalues that training leaves at zero. The machine trained at $k_{\rm tr}=\caKmin$ reaches an error of \caShErrFar{} at $k=1.57$.

For the acoustic branch, the two machines trained at small $k_{\rm tr}$ follow the same line above a floor of about \caAcErr{}, the residual at $k_{\rm tr}$. The machines trained at larger $k_{\rm tr}$ scatter about it because their acoustic fit already includes finite-wavenumber terms. Isolated points far below the line occur where the signed error changes sign. There, the residual at $k_{\rm tr}$ is canceled by the growing finite-wavenumber contribution. The damping is accurate by cancellation at those points, without comparable accuracy over a neighboring range. The data thus constrain the hydrodynamic branches where they are sampled.

\subsection*{S4. The graph machine}

The graph machine in \textit{Three properties tested on a graph} in the main text is a phase-space network with the three properties imposed on a road-like graph and trained on a traffic process. This section describes the process, learners and error measures, then derives the exact flux law. It gives the training results, spectra, slow-manifold test and closed-law comparison, followed by the three ablations.

\paragraph{Process, learners and measures.}
The graph is constructed from a Delaunay graph of 60 seed points in the unit square, with edges removed to limit the degree to 8. Capacities in each direction are drawn uniformly from $(0.05,0.25)$. The flow from $v$ to $w$ is $q_{vw}=\chi_{vw}\rho_v(1-\rho_w)$, clipped so that no node sends more than it holds. The data consist of 200 trajectories of 120 steps, starting from counts drawn uniformly from $(0,0.8)$ and extended to 300 steps for evaluation. Of these trajectories, 160 are used for training. The process relaxes toward a fixed point, with the rms deviation from the final state falling from 0.24 at $t=0$ to 0.012 at $t=30$. Long-time errors therefore measure the approach to that point rather than sustained dynamics.

The machine's target perceptron has two hidden layers of width 32 and a softmax allocation. The relaxation rate is learned as $\gamma=2\,\mathrm{sigmoid}(\eta)$, where $\eta$ is a trainable scalar initialized so that $\gamma=1.5$. The recurrent message-passing network has the same state size and width, with one $\tanh$ layer per message and capacities included in the input. Both learners use Adam with a learning rate of $2\times10^{-3}$ and cosine decay over 4000 steps. They are trained in batches of 16 windows of 30 steps, using the same window sequence.

Each training window starts at a step drawn uniformly between 0 and 90 of its trajectory. About two thirds of the windows therefore lie in the relaxed part of the approach to the fixed point. The learners are tested largely near equilibrium, and the no-change reference measures how much dynamics remains. Every window and rollout starts with the counts in the rest slot and empty edge slots. No state is carried between windows. The error $e_\rho$ is the rms count error over nodes, normalized by the rms of the data and averaged over the 40 held-out trajectories. The error $e_{\rm law}$ uses the same measure for the difference between the counts of the law and those of the machine.

Each ablation changes one construction element in the same phase-space network, and the network is retrained on the same data, with the state, shift, width, optimizer, windows and budget unchanged. The conservation ablation frees the allocation from summing to the count. The gap ablation fixes $\gamma$ at $2.0$ or $2.3$, and the projector ablation allows the target to depend on the allocation. In this construction, the projector property also requires conservation (section S1), so removing conservation also removes idempotency. At the uniform counts $\rho_v=0.4$, the rms over slots of $\Eq(\Eq(z))-\Eq(z)$ is 7\% of that of $\Eq(z)$. The first and third ablations therefore distinguish conservation from dependence of the target on the counts alone.

\paragraph{The exact flux law.}
For an undirected edge between $v$ and $w$, let $z^*_t(v,r_{vw})$ and $z^*_t(w,r_{wv})$ denote the post-collision values of $z(v,r_{vw})$ and $z(w,r_{wv})$. The shift places $z^*_t(v,r_{vw})$ in $z(w,r_{wv})$ and $z^*_t(w,r_{wv})$ in $z(v,r_{vw})$. The targets for the next collision are evaluated at the counts of step $t+1$. The two post-collision populations then satisfy
\begin{equation}
z^*_{t+1}(v,r_{vw})=(1-\gamma)\,z^*_t(w,r_{wv})+\gamma\,\Eq(v,r_{vw}),\qquad z^*_{t+1}(w,r_{wv})=(1-\gamma)\,z^*_t(v,r_{vw})+\gamma\,\Eq(w,r_{wv}).
\end{equation}
The net post-collision flux is $J_{vw}=z^*_t(v,r_{vw})-z^*_t(w,r_{wv})$. The count at $v$ decreases by the net outgoing flux, so the counts and fluxes obey the exact law
\begin{equation}
\label{eq:graphflux}
\rho_v(t+1)=\rho_v(t)-\sum_w J_{vw}(t),\qquad
J_{vw}(t)=\gamma\bigl[\Eq(v,r_{vw})-\Eq(w,r_{wv})\bigr](\rho(t))-(1-\gamma)\,J_{vw}(t-1).
\end{equation}
This law is closed in the counts and one kinetic variable per edge. Starting from empty edge slots gives $J_{vw}(-1)=0$, and the law reproduces the machine to rounding precision for every $\gamma$ tested. Eliminating the flux gives an exact equation for the counts alone with one step of memory,
\begin{equation}
\label{eq:graphcounts}
\rho_v(t+1)=\gamma\rho_v(t)+(1-\gamma)\rho_v(t-1)-\gamma\sum_w\bigl[\Eq(v,r_{vw})-\Eq(w,r_{wv})\bigr](\rho(t)),
\end{equation}
with $\rho_v(-1)=\rho_v(0)$. The symmetric combination $z^*_t(v,r_{vw})+z^*_t(w,r_{wv})$ is multiplied by $1-\gamma$ at each step and driven by the sum of its two targets. It does not enter the count change. The rest variable $z(v,0)$ is not shifted. Its deviation from the target is also multiplied by $1-\gamma$ at each step and forced by changes in the target between steps. It enters the count change only through the counts themselves. On the slow count modes, $J_{vw}$ changes little between steps. Setting $J_{vw}(t)=J_{vw}(t-1)$ in \eqref{eq:graphflux} then gives the first-order approximation \eqmain{12}.

\begin{figure}[!t]%
\centering
\includegraphics[width=\linewidth]{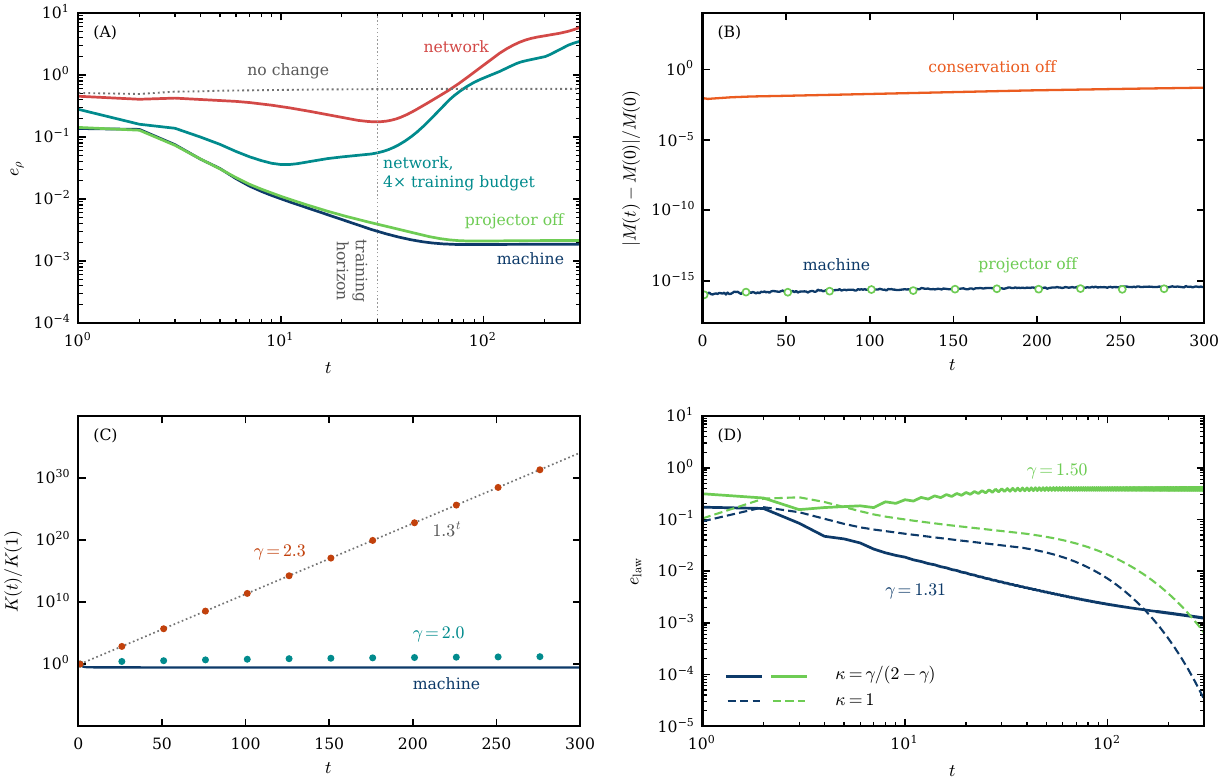}
\caption{Ablations of the graph machine. (A) Held-out error $e_\rho$ against rollout length for the graph machine, the message-passing network at two training budgets, the projector ablation and the no-change reference (dotted). The vertical line marks the training horizon. (B) Relative change in total count against time for the machine and the networks trained without conservation and without the projector. (C) Kinetic part $K(t)$ relative to its value at $t=1$ for the machine and the networks with $\gamma$ fixed at $2.0$ and $2.3$ (dots), with the growth $1.3^t$ (dotted). (D) Error $e_{\rm law}$ of the closed law in the counts alone relative to the machine, for learned $\gamma$ and for $\gamma$ fixed at $1.5$, with $\kappa=1$ (dashed) and $\kappa=\gamma/(2-\gamma)$ (solid).}
\label{fig:S4}
\end{figure}

\paragraph{Training and spectra.}
The graph machine and the message-passing network have $\ccParMachine$ and $\ccParBase$ parameters, respectively. The machine reaches a window loss of \ccLossMachine{}, while the network reaches \ccLossBase, a hundred times larger. Increasing the network's training budget fourfold lowers its loss by a factor of eight, but its rollouts still diverge (Fig.~\ref{fig:S4}(A)). Its total count drifts by up to $\ccMassB\%$ over the rollout.

The machine's one-step Jacobian, linearized about the uniform state $\rho_v=0.4$, has $\ccJacDim$ eigenvalues. Of these, $\ccJacKinCount$ have modulus $|1-\gamma|=\ccOneMinusGamma$, and about a hundred lie near $\sqrt{|1-\gamma|}$ as complex pairs coupling a count to an edge flux. Seven non-conserved eigenvalues have modulus above $0.6$. The largest non-conserved eigenvalue at this state, $\lambda_1=\ccJacMachine$, corresponds to the slowest count mode. The gap of the composed step thus separates a kinetic cluster set by relaxation from a small number of slow count modes. The message-passing network's one-step Jacobian has $\ccJacBaseAbove$ eigenvalues of modulus above one.

\paragraph{The slow manifold.}
A perturbation of the message-passing network's hidden channels that initially leaves its readout unchanged grows by a factor of $\ccBaseHiddenGrowth$ at $t=300$. Its imprint on the counts reaches $\ccBaseImprintHundred\%$ at $t=100$, showing that the hidden state retains memory of the perturbation. For the graph machine, the imprint of a different allocation on the counts is $\ccImprintOne\%$ of their rms after the first step and \ccImprintHundred{} of their rms at $t=100$.

\paragraph{The closed law.}
The slowest eigenvalue of the count dynamics provides a comparison without fitted parameters. Its value is $\ccJacMachine$ for the machine. The law with $\kappa=\ccFluxFactor$, determined by the learned $\gamma=\ccGamma$, gives $\ccJacRenorm$. Setting $\kappa=1$, as for a direction-preserving shift, instead gives $\ccJacClosed$. A machine trained with $\gamma$ fixed at $0.5$ confirms the coefficient $\kappa=\gamma/(2-\gamma)$ at a rate below one. The slowest count eigenvalue is then $\ccJacRenormHalf$ for the law, compared with $\ccJacMachineHalf$ for the machine.

For the machine with learned $\gamma$, the error $e_{\rm law}$ of the slaved law along the held-out trajectories has a maximum of 1.9\% after the first ten steps. At $\gamma=1.5$, the coefficient is $\kappa=3$, and the explicit one-step law in the counts alone is unstable away from uniform states (Fig.~\ref{fig:S4}(D)). The machine itself, which retains $J$ as a variable, remains stable in this test. The first-order form of the emergent law therefore has its own stability limit at large coefficient.

Only the divergence of the flux is identifiable from the counts. The negative node-level divergence of the slaved learned flux has a correlation of $0.99$ with the one-step count change of the process, while the net edge fluxes do not correlate. Flows around graph cycles leave the counts unchanged.

\paragraph{The ablations.}
Without conservation, the allocation sums to $\ccAbOneSumMean$ on average over the data states, but its rms deviation node by node is $\ccAbOneSumRms\%$. The total count changes by up to $\ccAbOneMassThree\%$ at $t=\ccTlong$ (Fig.~\ref{fig:S4}(B)), and the held-out error rises to $\ccAbOneErrThree\%$. The reduction to counts and fluxes differs from the network by \ccAbOneClosure\% because it has no source term. With the learned source $\gamma[\Eq(v,0)+\sum_r\Eq(v,r)-\rho_v]$ added to the count balance, it reproduces the network to rounding precision. The one-step Jacobian also loses its eigenvalue at one. The eigenvalue closest to one is $\ccAbOneJacMax$ at the probe state.

With $\gamma$ fixed at $2.0$ to remove the gap, $\ccAbTwoUnitCircle$ of the $\ccAbTwoJacDim$ eigenvalues of the one-step Jacobian lie on the unit circle. The net flux then has eigenvalue $\gamma-1=1$, so it accumulates flux differences instead of forgetting them. The counts follow \eqref{eq:graphcounts} without damping, and the trained network is weakly unstable. The exact law in $(\rho,J)$ still holds to rounding precision.

To test a first-order law in the counts alone, we fit a softmax-flux network of the machine's form by regression on 40 rollouts of the machine. The fit uses 1500 optimization steps and 30-step windows starting between steps 2 and 90. The fitted law is rolled out from the counts at step 2. It differs from the held-out rollouts by $\ccAbTwoFitLaw\%$ at both 30 and 100 steps. The machine's held-out error against the process is $\ccAbTwoErrThreeTwo\%$.

With $\gamma$ fixed at $2.3$, the kinetic part grows by a factor of $\ccAbTwoKinGrowth$ per step (Fig.~\ref{fig:S4}(C)). Every non-conserved eigenvalue has modulus above one, and the rollouts diverge. Training lowers the window loss by a factor of thirty, but it remains $\ccAbTwoLossThreeOverNoChange$ times the loss of predicting no change.

Without the projector, the error at $t=\ccTlong$ is $\ccAbThreeErrThree\%$, compared with $\ccErrMThree\%$ for the machine (Fig.~\ref{fig:S4}(A)).

\renewcommand{\refname}{Supplementary References}

\end{document}